\documentclass[aps,prd,onecolumn,nofootinbib,superscriptaddress]{revtex4-1}

\usepackage{amsmath,amssymb,slashed}
\usepackage{dsfont}
\usepackage{graphicx}
\usepackage{epstopdf}  
\usepackage{slashed}
\usepackage{hyperref}
\usepackage{xcolor}

\begin{document}

\title{Electromagnetic form factors of the transition from the Delta to the nucleon}

\newcommand{\UU}{Institutionen f\"or fysik och astronomi, Uppsala universitet, Box 516, S-75120 Uppsala, Sweden}
\newcommand{\SUT}{School of Physics, Suranaree University of Technology, 111 University Avenue, Nakhon Ratchasima 30000, Thailand}
\newcommand{\Bolder}{University of Colorado Boulder, Department of Electrical, Computer, and Energy Engineering, Boulder, USA}

\author{Moh Moh Aung} \affiliation{\SUT}\affiliation{\UU}
\author{Stefan Leupold} \affiliation{\UU}
\author{Elisabetta Perotti} \affiliation{\UU} \affiliation{\Bolder}
\author{Yupeng Yan} \affiliation{\SUT}

\date{\today}

\begin{abstract}
The low-energy electromagnetic form factors of the  $\Delta$(1232)-to-nucleon transition are derived combining dispersion theory techniques and chiral perturbation theory. The form factors are expressed in terms of the well-understood pion vector form factor and pion-baryon scattering amplitudes. 
Nucleon and Delta exchange terms and contact terms constitute the input for these pion-baryon amplitudes. The framework is formulated for all form factors. When comparing to experimental data in the spacelike region of $e^- N \to e^- \Delta$ scattering, the focus lies on the numerically dominant magnetic dipole transition form factor. Fitting two subtraction constants (one for the scattering amplitude, one for the form factor) yields a very good description of this dominant form factor up to photon virtualities of about 0.6 GeV. After determining the subtraction constants in the spacelike region and at the photon point, respectively, predictions for the timelike region of Dalitz decays $\Delta \to N \, e^+ e^-$ are presented. 
\end{abstract}

\maketitle

\section{Introduction and summary}
\label{sec:intro}
One of the interesting features of quantum field theory is the possibility that electrically neutral particles can interact with photons. Even the Higgs boson as an ``elementary'' particle couples to two photons \cite{Aad:2012tfa,Chatrchyan:2012ufa}. In turn this means that one can use electromagnetic processes to learn something about uncharged objects. Concerning the neutron, its exploration went through several levels of understanding of its properties and structure, and this process has not come to an end yet. The neutron's non-vanishing magnetic moment points to its substructure in terms of electrically charged objects, the quarks. Yet, the example of the two-photon decay of the Higgs boson tells that charged constituents are not the only possibility to explain a non-trivial electromagnetic response of an object described by quantum field theory. Also virtual many-body fluctuations can significantly contribute. Indeed the spatial charge distribution of the neutron \cite{Miller:2007uy} with a negatively charged surface suggests fluctuations into a negative pion and a proton. Since the pion is so much lighter than other hadrons, a phenomenon caused by spontaneous chiral symmetry breaking, the large-distance electromagnetic response of the neutron is dominated by pion physics, irrespective of the detailed quark substructure. What really points to the composite nature of the neutron (and the proton) is the possibility to excite it to a higher-lying state. And also here electromagnetic processes come into play and pion physics is important. 
In this work we address the excitation of the neutron to the $\Delta^0$ resonance. Of course, this process is related to the excitation of the proton to $\Delta^+$ via isospin symmetry, but we regard the electromagnetic excitation of a neutral object as more intriguing. Therefore we will formulate many aspects of our framework in terms of neutrons and neutral Deltas. Yet, when comparing to data we utilize isospin symmetry and compare to the better measured reactions with protons and positive Delta states.

\begin{figure}[h!]
\centering
    \includegraphics[width=\textwidth]{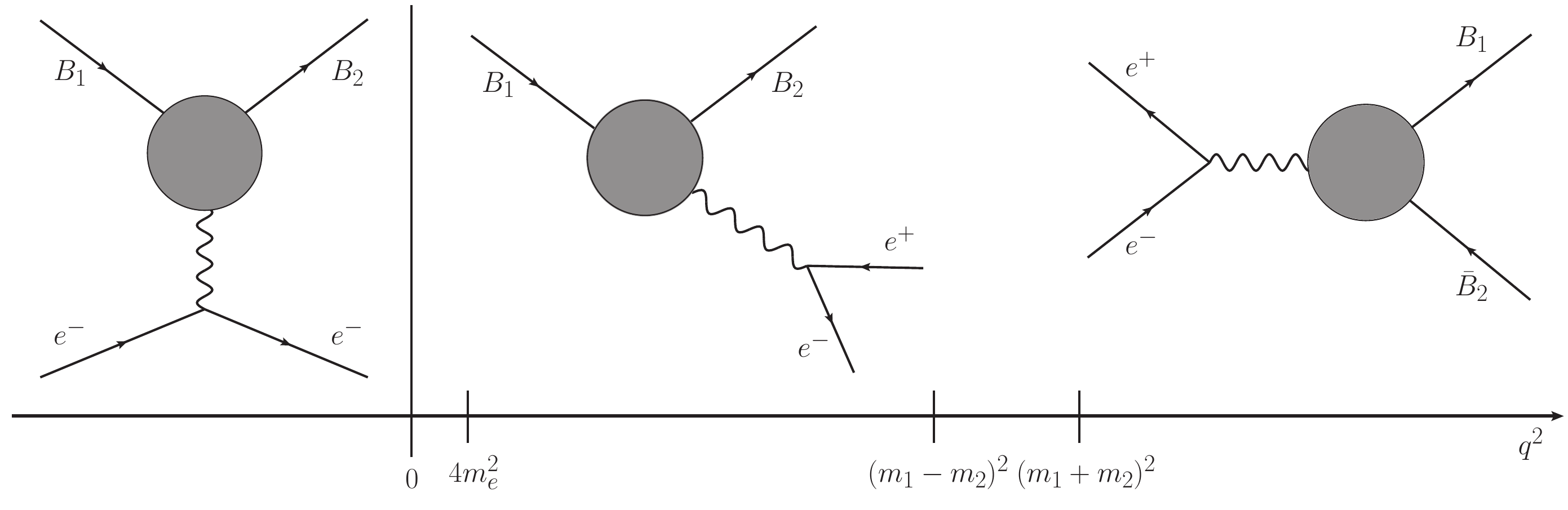}
    	\caption{Space- and timelike FFs can be accessed exploiting different reactions. Processes proceed from left to right. The momentum of the virtual photon (wavy line) is denoted by $q$, the mass of electron $e^-$ and positron $e^+$ by $m_e$, the mass of baryon $B_i$ by $m_i$, $i=1,2$.}
    	\label{fig:FFq2}
\end{figure}

Electromagnetic form factors (FFs) allow for a general characterization of composite objects, even when the intrinsic structure of said objects is not yet fully understood. The FFs are functions of the momentum transfer squared $q^2$, which means that they can be experimentally addressed in different kinematical regions, based on the choice of the reaction.  Fig.\ \ref{fig:FFq2} displays both the spacelike $(q^2<0)$ and the timelike $(q^2>0)$ regions. By scattering electrons on a baryon, $e^-\,B_1\rightarrow  e^-\,B_2 $,  one has access to the spacelike region. The timelike region is explored via Dalitz decays  into a lighter baryon and an electron-positron pair, $B_1\rightarrow B_2 \, e^+e^-$,  and via electron-positron annihilation into a baryon-antibaryon pair, $e^+e^-\rightarrow B_1 \, \bar{B}_2$. 
The electromagnetic FFs  are useful tools to investigate the intrinsic properties of  hadrons. The interest in their determination is driven by the desire to better understand the properties of matter \cite{Punjabi:2015bba,Pascalutsa:2006up}. 

This work focuses on the transition from the Delta to the nucleon and is the third in a series of similar studies. The Uppsala (UU) group has previously considered other transition form factors (TFFs), namely  the $\Sigma^0$-$\Lambda$ TFFs  \cite{Granados:2017cib} (which are the only ones among the ground-state spin-1/2 baryons), and the $\Sigma^{*0}$-$\Lambda$ TFFs (which involve a decuplet baryon) \cite{Junker:2019vvy}. These, as well as the $\Delta^0$-$n$ TFFs, have in common that they are purely isovector FFs. 
However, the experimental situation is rather different for these TFFs. 
The unstable nature of hyperons and therefore the major complication to obtain spacelike TFFs from fixed-target experiments\footnote{One would need a hyperon beam that scatters on the electron cloud of atoms in the target.} motivated our previous research program where we explored the possibility to learn about the spacelike region from dispersion theory and input from Dalitz decay measurements in the timelike region (see also \cite{Husek:2019wmt}).
In fact, the spacelike region is of primary interest from the point of view of hadron-structure studies. It is where the interpretation of the FFs as spatial distributions of e.g.\ electric charge is possible \cite{Miller:2007uy,Tiator:2011pw,Alarcon:2017asr}.

In the present work we turn from hyperons to those low-lying baryons where the minimal quark content is provided solely by the up and down quarks. These are the nucleon and the $\Delta(1232)$. Such studies will serve to scrutinize our methods developed in the hyperon sector. The direct vector-isovector FFs of the nucleon have been addressed already in \cite{Leupold:2017ngs,Alvarado:2023loi}. Here we study the $\Delta$-nucleon transitions. Of course, such studies are also interesting in their own right, irrespective of the exploration potential in the hyperon sector. Recall that the previous strategy has been to provide predictions for the low-energy spacelike TFFs of hyperons, for which there are no experimental data. This can be achieved provided that the TFFs are measured at least in the low-energy timelike region, and by means of theoretical tools like dispersion relations and chiral perturbation theory ($\chi$PT).
Thanks to dispersion relations, the TFFs can be analytically continued into the experimentally not easily accessible region. 

Rather opposite to the hyperon case, there are data for the $\Delta$-nucleon TFFs in the spacelike region, see \cite{Pascalutsa:2006up,OOPS:2004kai,Sparveris:2006uk,Stave:2006ea,CLAS:2009ces} and further references therein. This fact makes us turn our strategy around compared to the hyperon case. Thus the scope of this paper is twofold. On the one hand, we want to make predictions for the $\Delta^0$-$n$ TFFs (coinciding with $\Delta^+$-$p$ in the isospin limit) and in particular for Dalitz decays. On the other hand, we want to test the validity of our methods against real data. 

Our framework \cite{Granados:2017cib,Leupold:2017ngs,Junker:2019vvy,Alvarado:2023loi,AnDi-prep} is based on dispersion theory, a model-independent approach justified by the fundamental properties of local relativistic quantum field theory: micro-causality, analyticity, unitarity and crossing symmetry. Restricting ourselves to low energies allows for an effective-field-theory point of view. Only the low-lying degrees of freedom need to be considered explicitly, the short-distance physics is covered by contact terms (subtraction constants in the language of dispersion relations). The challenge is to include the longer-distance physics in a model-independent way. Effectively our framework includes the physics of pions, rho mesons, nucleons and Deltas. In principle, $\chi$PT \cite{Weinberg:1978kz,Gasser:1983yg,Gasser:1984gg,Scherer:2002tk,Scherer:2012xha,Hilt:2017iup,Unal:2021byi}, the model-independent low-energy incarnation of QCD, includes nucleons and pions and allows for an extension to $\Delta$ states. But it is difficult to include the rho meson in a model-independent way; see also the corresponding discussions in \cite{Terschlusen:2016kje,Alvarado:2023loi}. On the other hand, the rho meson is an elastic resonance of pion-pion scattering. The pion phase shifts are very well known \cite{Colangelo:2001df,Garcia-Martin:2011iqs}. Dispersion theory allows for the systematic inclusion of phase shifts. In this way, the physics of the rho meson is covered in a model-independent fashion. Concerning electromagnetic FFs at low energies, the virtual photon couples dominantly to the lightest degrees of freedom, the pions. This is quantified by the pion vector form factor, which is also very well known \cite{Hanhart:2012wi}. The additional ingredient are pion-baryon scattering amplitudes. Since the physics of the rho meson is already covered by the dispersive part, the pion-baryon scattering amplitudes need to cover the physics of the lightest baryon degrees of freedom. For the hyperon cases \cite{Granados:2017cib,Junker:2019vvy}, the UU group used $\chi$PT to obtain those hadronic amplitudes because there are no direct data on pion-hyperon scattering. We follow here the same path. 

In this work we use dispersive representations for the TFFs or linear combinations thereof. A comparison to previous works is in order here. The reaction $e^- N \to e^- \Delta$ is part of the physical reaction $e^- N \to e^- \, N \pi$. Somewhat more formally this is $e^- \to e^- \gamma^*$ with subsequent $\gamma^* N \to \Delta \to N \pi$. Here the $\Delta$ is typically extracted from a partial-wave analysis since the $N\pi$ system does not only couple to the $\Delta$ resonance. Dispersion theory for the $N\pi$ system has been used among other methods; for a review see \cite{Pascalutsa:2006up}. In this way, TFFs have been extracted from data. But dispersion theory has not been used for the TFFs themselves, i.e.\ it has not been used for the $\gamma^*$ invariant mass instead of the $N\pi$ invariant mass. We fill this gap with our present exploratory study of the use of dispersion relations for the FFs of the Delta-to-nucleon transition. 

We regard it also as important to stress what will {\em not} be covered by our present work. We do not aim at a complete description of the (formal) reaction $\gamma^* N \to \Delta \to N \pi$. Instead we focus on the photon virtuality. This implies in particular that we treat the $\Delta$ as if it were an asymptotic state with the peak mass of the resonance. A fully dispersive treatment of both virtualities of $\gamma^*$ and $N\pi$ is way beyond the scope of the present work. We note however that other works (using $\chi$PT) deal with the $\Delta$ as a pole in the complex plane \cite{Hilt:2017iup,Unal:2021byi} and not as a stable state. Such a more realistic treatment in dispersion theory instead of $\chi$PT would require a generalization of the optical theorem to the complex plane, which is also beyond the scope of the present work.

In addition, we note that one crucial input of our formalism are the scattering amplitudes $\Delta$-$\pi$ to $N$-$\pi$. In line with our previous works \cite{Granados:2017cib,Leupold:2017ngs,Junker:2019vvy,Alvarado:2023loi}, we determine these amplitudes from $\chi$PT and unitarize them in the Muskhelishvili-Omn\`es framework. This means that in the formal cross channel $\Delta$-$\bar N$ to $\pi$-$\pi$ the pion rescattering is fully taken into account by the pertinent pion phase shift, i.e.\ we have a framework for the formal reaction $\Delta \bar N \to \pi \pi \to \pi \pi$ that respects unitarity and analyticity. The only required input is the baryon dynamics. Following \cite{Granados:2017cib,Leupold:2017ngs,Junker:2019vvy,Alvarado:2023loi}, we cover this aspect by $\chi$PT. This involves diagrams with an intermediate $\Delta$ propagator corresponding to the formal scattering reaction $\Delta \pi \to \Delta \to N\pi$. Given that the $\Delta$ is an elastic resonance in the $N$-$\pi$ channel, those $\Delta$ exchange diagrams might be seen as an approximation to the full rescattering into $N$-$\pi$, formally $\Delta \pi \to N \pi \to N\pi$. This means that we treat $N \pi$ intermediate states in the baryon-pion channels differently from $\pi \pi$ intermediate states in the crossed channel. Since the elastic nucleon-pion scattering amplitudes are very well known \cite{Hoferichter:2015hva}, there would be an alternative to our treatment of the scattering amplitudes in $\chi$PT. In principle, one could deal with such scattering amplitudes in the very same way concerning direct and crossed channels. For instance, such a framework has been formulated for the amplitudes $\eta'$-$\pi$ to $\eta$-$\pi$ in \cite{Isken:2017dkw}. Given that we treat the pion-pion rescattering dispersively, it would be appealing to treat also the pion-nucleon rescattering dispersively. On the other hand, we want to test the quality of our hyperon calculations \cite{Granados:2017cib,Junker:2019vvy} where no data exist for hyperon-pion scattering amplitudes. But it should be clear that our input, $\chi$PT (at next-to-leading order), is an approximation. Therefore one should not expect that we can obtain an accuracy at the percent level.

In view of all these possible future improvements --- complete description of the reaction $\gamma^* N \to N \pi$ and/or treating the $\Delta$ as a pole in the complex plane and/or treating not only pion-pion scattering but also pion-nucleon scattering dispersively --- we regard the present work as a first exploratory study of our formalism. Consequently we will develop the formalism for all TFFs but focus for the results on the dominant (magnetic) TFF. We will see that we obtain very encouraging results for the magnetic dipole TFF. With two subtraction constants --- one for the TFF and one for the baryon-meson scattering amplitude --- fitted to the spacelike data, we can reproduce the magnetic TFF up to photon virtualities of about 0.6 GeV in the spacelike region. The subtraction constants parametrize our ignorance of the short-distance physics while the longer-range physics (pion rescattering in the 2-pion channel and nucleon and Delta exchange in the baryon-pion channels) is explicitly covered by our framework. The results are practically insensitive to the phenomenological uncertainties of our other input parameters that are related to the longer-range physics and not fitted to the TFF data but obtained from other sources. The successful description of the spacelike data suggests that the previous hyperon calculations have also a solid foundation and can be used to translate experimental information from the timelike to the spacelike region as proposed, for instance, in \cite{Granados:2017cib}. 

At low energies, the other two TFFs (electric and Coulomb quadrupole) are much smaller than the magnetic TFF (few-percent level) \cite{Pascalutsa:2006up}. In view of our approximations and the exploratory nature of our present work, we do not expect that we can reproduce these smaller TFFs very well. Still we will address the electric TFF in an extended outlook section and find that also these results are encouraging. Yet the main focus of the present work will lie on the magnetic transition. 

The rest of the paper is structured in the following way. In Section \ref{sec:tffobs}, we introduce first TFFs that are free of kinematical constraints. We relate them to helicity amplitudes and to Jones-Scadron TFFs \cite{Jones:1972ky}. TFF ratios commonly used by the experimental groups are introduced for the spacelike region of $\Delta$ production in electron-nucleon scattering. The TFFs are also related to differential decay rates in the timelike region of Dalitz decays $\Delta \to N \, e^+ e^-$. In Section \ref{sec:disp}, we recall the main aspects of the dispersive framework that has been provided already in \cite{Junker:2019vvy}. One interesting aspect is the appearance of an anomalous cut. An appendix is reserved for the technical aspects. In practice, we use the dispersion relations for Jones-Scadron TFFs, in particular for the dominant magnetic dipole TFF. In Section \ref{sec:chiPT}, we specify the input from $\chi$PT at leading and next-to-leading order, used for the baryon-pion scattering amplitudes that enter the dispersive framework. In Section \ref{sec:results} we discuss in detail the results for the magnetic TFF and related quantities. We provide predictions for Dalitz decay distributions. Section \ref{sec:outlook} constitutes an extended outlook where we explore whether our formalism is accurate enough to address one of the smaller Jones-Scadron TFFs, namely the electric quadrupole TFF. Several appendices are added where we discuss technical aspects and details that would interrupt too much the central line of reasoning in the main text.

\section{Transition form factors and observables}
\label{sec:tffobs}

The three $\Delta$-$n$ TFFs are defined in agreement with \cite{Korner:1976hv,Junker:2019vvy,Salone:2021bvx} as
\begin{equation}
  \label{eq:defTFF}
  \langle 0 \vert j_\mu \vert \Delta  \bar n \rangle 
  = e \, \bar v_n(p_n,\lambda) \, \Gamma_{\mu\nu}(p_{\Delta},p_{n}) \, u_{\Delta}^\nu(p_{\Delta},\sigma) 
\end{equation}
with
\begin{eqnarray}
  \Gamma^{\mu\nu}(p_{\Delta},p_{n}) 
  & := & -(\gamma^\mu q^\nu - \not \!q \, g^{\mu\nu}) \, m_{\Delta} \, \gamma_5 \, F_1(q^2)   \nonumber \\ && {}
  + (p_{\Delta}^\mu q^\nu - p_{\Delta} \cdot q \, g^{\mu\nu}) \, \gamma_5 \, F_2(q^2)   \nonumber \\ && {}
  + (q^\mu q^\nu - q^2 \, g^{\mu\nu}) \, \gamma_5 \, F_3(q^2) 
  \label{eq:defTFF2}
\end{eqnarray}
and $q:=p_{\Delta}+p_n$. The neutral spin-3/2 Delta hyperon is denoted by $\Delta$. Conventions for the spin-3/2 spinor $u^\mu$ can be found in the Appendix B of Ref.\ \cite{Junker:2019vvy}. 
The helicities of $\Delta$ and $\bar n$ are denoted by $\sigma$ and $\lambda$, respectively.
The TFF definition of Eq.\ \eqref{eq:defTFF} reflects the fact that from a formal point of view we are interested in the process $\Delta \bar n \rightarrow \pi^+\pi^- \rightarrow \gamma^*$. This is what enters our dispersive calculation. In practice, however, other kinematical regions such as $e^+e^-\rightarrow \gamma^*\rightarrow {\Delta} \bar n$, $e^- n\rightarrow  e^-\Delta$ and $\Delta\rightarrow n\gamma^*\rightarrow n \, e^+e^-$ can be experimentally studied. In addition, the very same TFFs enter $e^+e^-\rightarrow \gamma^*\rightarrow \Delta^+ \bar p$, $e^- p \rightarrow  e^-\Delta^+$ and $\Delta^+\rightarrow p\gamma^*\rightarrow p \, e^+e^-$, respectively. 

For the amplitude relevant for the Dalitz decay, $\Delta \to n \, \gamma^*$, one finds
\begin{equation}
  \label{eq:defTFFDalitz}
  \langle n \vert j_\mu \vert \Delta \rangle 
  = e \, \bar u_n(p_n,\lambda) \, \Gamma_{\mu\nu}(p_{\Delta},-p_{n}) \, u_{\Delta}^\nu(p_{\Delta},\sigma) \,.
\end{equation}
This leads to the very same expression as on the right-hand side of \eqref{eq:defTFF2} but with 
$q:=p_{\Delta}-p_n$. 

For later use we introduce the K\"all\'en function
\begin{eqnarray}
  \label{eq:kallenfunc}
  \lambda(a,b,c):=a^2+b^2+c^2-2(ab+bc+ac) \,,
\end{eqnarray}
which appears frequently in the context of the kinematics of a two-body decay.

We construct linear combinations of $F_1$, $F_2$ and $F_3$, which correspond to TFFs with fixed helicity combinations.
These TFFs are denoted by $G_m$, with $m=\sigma-\lambda=0,\pm 1$, and are given by 
\begin{eqnarray}
  G_{-1}(q^2) &:=&  (-m_n (m_n+m_{\Delta}) + q^2) \, F_1(q^2)  \nonumber \\ && {}
  +\frac12 \, (m_{\Delta}^2-m_n^2+q^2) \, F_2(q^2) + q^2 \, F_3(q^2) \nonumber \\ &&  \mbox{for}
  \quad \sigma=-\frac12 \,, \lambda=+\frac12  \,,
  \label{eq:F-1def}
\end{eqnarray}
\begin{eqnarray}
  G_{0}(q^2) &:=& m^2_{\Delta} \, F_1(q^2) + m^2_{\Delta} \, F_2(q^2) \nonumber \\ && {}
  + \frac12 \, (m_{\Delta}^2-m_n^2+q^2) \, F_3(q^2)   \nonumber \\ &&  \mbox{for}
  \quad \sigma=+\frac12 \,, \lambda=+\frac12 \,,
  \label{eq:F0def}
\end{eqnarray}
and 
\begin{eqnarray}
  G_{+1}(q^2) &:=& m_{\Delta} (m_n+m_{\Delta}) \, F_1(q^2) \nonumber \\ && {}
  + \frac12 \, (m_{\Delta}^2-m_n^2+q^2) \, F_2(q^2) + q^2 \, F_3(q^2)   \nonumber \\ &&   \mbox{for}
  \quad \sigma=+\frac32 \,, \lambda=+\frac12 \,.
  \label{eq:F+1def}
\end{eqnarray}

On the one hand, a more straightforward study of the pion-loop contributions to the TFFs  can be performed using the fixed helicity combinations (\ref{eq:F-1def}), (\ref{eq:F0def})  and (\ref{eq:F+1def}). On the other hand,  these helicity amplitudes are not independent quantities but satisfy kinematical constraints. 
At the production threshold $q^2=(m_n+m_{\Delta})^2$, one has $G_{+1}=G_{-1}=(m_{\Delta}+m_n) G_0/m_{\Delta}$. At the decay threshold $q^2 = (m_\Delta-m_n)^2$, one finds
\begin{equation}
    G_{+1}+G_{-1}=2(m_{\Delta}-m_n) \frac{G_0}{m_{\Delta}}  \,.
    \label{eq:kinconstr00}
\end{equation}
See also \cite{Ramalho:2016zzo,Salone:2021bvx} for further discussions. Those kinematical constraints can complicate the formulation of dispersion relations \cite{Bardeen:1969aw,Tarrach:1975tu}. 
Nonetheless, we will follow the approach of \cite{Junker:2019vvy} and formulate dispersion relations for helicity amplitudes. In practice, the constraints at the production threshold are far away from the low-energy region we are interested in. The constraint \eqref{eq:kinconstr00}, however, is relevant. We will frequently come back to this issue. 

In the literature one finds different conventions for the TFFs. In addition, there is a sign ambiguity related to the un-measurable overall phase of a quantum state $\vert \Delta \rangle$. In \cite{Carlson:1985mm}, where the transition from nucleon to $\Delta$ is considered, Carlson introduces TFFs  (here labeled with ``Ca'') which are related to our 
TFFs by\footnote{There is a mismatch between the conventions used in \cite{Carlson:1985mm} and here. This is 
essentially based on the fact that we introduce our TFFs via the coupling of a virtual timelike photon to a spin-3/2 
baryon and a spin-1/2 antibaryon where the latter has helicity $+1/2$. In \cite{Carlson:1985mm} the TFFs are introduced 
via the coupling of a virtual spacelike photon to an incoming spin-1/2 baryon and an outgoing spin-3/2 baryon. The former has 
helicity $+1/2$. If one translates our case to the one in \cite{Carlson:1985mm} our antibaryon turns to a baryon with 
helicity $-1/2$ and not $+1/2$. This sign change relates our TFF $G_m$ to Carlson's TFF $G^{\rm Ca}_{-m}$ 
for all $m=0,\pm 1$.}
\begin{eqnarray}
  G_-^{\rm Ca} &=& \frac{\zeta Q_-}{2 \, m_n} \, G_{+1}  \,, \nonumber \\
  G_+^{\rm Ca} &=& \frac{\zeta Q_-}{2 \sqrt{3} \, m_n} \, G_{-1}  \,, \nonumber \\
  G_0^{\rm Ca} &=& \frac{\zeta Q \, Q_-}{\sqrt{6} \, m_n \, m_{\Delta}} \, G_0  
  \label{eq:rel-Carlson}
\end{eqnarray}
with $Q_-:=\sqrt{Q^2+(m_n -m_{\Delta})^2}$.  Since one studies now reactions with $q^2 <0$, it is convenient to introduce $Q^2:=-q^2 > 0$ and $Q:=\sqrt{Q^2}$. We will come back to the factor $\zeta = \pm 1$ below.

In Ref.\ \cite{Pascalutsa:2006up} various conventions are related to each other, including the ones 
from \cite{Carlson:1985mm}. With the help of \eqref{eq:rel-Carlson} and \cite{Pascalutsa:2006up} our TFFs can be easily 
related to any other TFF combinations and conventions. 
In particular, we provide here also the relation to the Jones-Scadron TFFs \cite{Jones:1972ky} that are frequently used in the experimental analyses. 

It turns out that it is convenient to introduce linear combinations of the TFFs $G_m$. This allows to single out a specific combination that is much larger than the other two. This has been confirmed by experiment and is also in line with quark-model considerations and with QCD in the limit of a large number of quark 
colors  \cite{Pascalutsa:2006up,Ramalho:2015qna}. Thus it makes sense to study first the dominant (and therefore best measured) structure 
with our dispersion relations. In a second step, one could look at the smaller quantities. We postpone a full discussion of these smaller quantities
to future work, cf.\ also the corresponding discussion in Section \ref{sec:intro}. But we provide here the dispersive formalism for all TFFs and explore also one of the smaller quantities to see what can be achieved with the present formalism.

The Jones-Scadron FFs are very well suited to separate the small quantities from the dominant contribution. In terms 
of our FFs, one finds 
\begin{eqnarray}
  G_M^* &=& \frac{\zeta}{\sqrt{6}} \, \frac{m_n}{m_n+m_\Delta} \left( G_{-1} - 3 G_{+1} \right)  \,, \nonumber \\
  \label{eq:Gs-jones-scadron}
  G_E^* &=& -\frac{\zeta}{\sqrt{6}} \, \frac{m_n}{m_n+m_\Delta} \left( G_{-1} + G_{+1} \right)  \,, \\
  G_C^* &=& -\frac{4 \zeta}{\sqrt{6}} \, 
  \frac{m_n}{m_n+m_\Delta}  G_0  \,.
\nonumber  
\end{eqnarray}
At low energies, the magnetic dipole TFF, $G_M^*$, is much larger than the other two quantities, the electric and the Coulomb quadrupole TFF \cite{Pascalutsa:2006up}. Thus the idea is to study first dispersion relations for $G_M^*(q^2)$ (instead of $G_{\pm 1}$, separately). 
After studying how well we can reproduce $G_M^*(-Q^2)$ in the spacelike region, we will provide predictions for the timelike region. Finally we will go through the same steps with $G_E^*(-Q^2)$ but reserve the discussion of  $G_C^*(-Q^2)$ for future work. 

Interestingly, the kinematical constraint \eqref{eq:kinconstr00} relates the electric and Coulomb TFFs; a relation expressed by the Siegert theorem \cite{Ramalho:2016zzo}:
\begin{eqnarray}
  \label{eq:constr-EC*}
  G_C^*((m_\Delta - m_N)^2) = \frac{2m_\Delta}{m_\Delta-m_N} \, G_E^*((m_\Delta - m_N)^2)  \,.
\end{eqnarray}
We will come back to this relation in the context of dispersion relations at the end of Section \ref{sec:disp}. 

When comparing to experimental results carried out in the spacelike region of electroproduction, there is a subtlety related to the fact that the TFFs of unstable resonances can be complex in the spacelike region \cite{Junker:2019vvy}. On the other hand, the experimental results constitute real-valued quantities \cite{Pascalutsa:2006up,OOPS:2004kai,Sparveris:2006uk,Stave:2006ea,CLAS:2009ces}. Thus we have to determine which quantities are actually meant by real-valued, published results denoted by $G^*_{M,E,C}$. As we will show in Appendix \ref{sec:exp-re-BW} it is actually the respective real part. 

To adjust to the conventions used for the pion electroproduction we note that \cite{Pascalutsa:2006up}
\begin{eqnarray}
  \label{eq:GM-iMM}
  {\rm Re}G_M^*(-Q^2) = \frac{8 m_n m_\Delta}{e \, (m_n+m_\Delta) \, Q_-} \, 
  \sqrt{\frac{2 \pi k_\Delta \Gamma_\Delta}{3}} \, {\rm Im}M_{1+}^{(3/2)} 
\end{eqnarray}
with $Q_-=\sqrt{Q^2+(m_\Delta-m_N)^2}$, the electric charge $e = \sqrt{4\pi \alpha} \approx 0.303$, 
the $\Delta$ decay width $\Gamma_\Delta \approx 0.117\,$GeV, 
and the momentum of the 
pion as produced in the $\Delta$ rest frame, $k_\Delta = \lambda^{1/2}(m_\Delta^2,m_n^2,m_\pi^2)/(2 m_\Delta) \approx 0.229\,$GeV. The pertinent pion electroproduction amplitude is denoted by $M_{1+}^{(3/2)}$. 

It is common practice on the experimental side to study ${\rm Im}M_{1+}^{(3/2)}$ and the multipole ratios $R_{EM}$ and $R_{SM}$ instead of the TFFs $G^*_{M,E,C}$. Those multipole ratios are given by \cite{Pascalutsa:2006up} 
\begin{eqnarray}
  R_{EM} &=& -\frac{{\rm Re}G^*_E}{{\rm Re}G^*_M} \,, \nonumber \\
  R_{SM} &=& - \frac{\lambda^{1/2}(-Q^2,m_\Delta^2,m_n^2)}{4 m_\Delta^2} \frac{{\rm Re}G^*_C}{{\rm Re}G^*_M}   \,.
    \label{eq:multip-ratios}
\end{eqnarray}

Finally we come back to the phase factor $\zeta$ that appears in \eqref{eq:rel-Carlson} and \eqref{eq:Gs-jones-scadron}. It has been introduced since the experimental groups might use a different convention when extracting their helicity amplitudes or TFFs (see, e.g., \cite{Tiator:2011pw}). In fact, when introducing a quantum field to represent the $\Delta$ one has an undetermined quantum mechanical phase. This freedom might be used to make a choice for $\zeta$ in (\ref{eq:Gs-jones-scadron}) {\em or} (exclusive ``or'') to make a choice for the sign of the $\Delta$-$N$-$\pi$ coupling constant $h_A$; see Section \ref{sec:chiPT} below. In the following we will choose 
\begin{equation}
    \zeta = -1  \,.
    \label{eq:zeta-choice}
\end{equation}
This means that we have to explore both sign options for the coupling constant $h_A$. We will do that in a two-step procedure. First, we will explore QCD for a large number of colors $N_c$ to get a first indication of which sign of $h_A$ fits to the choice (\ref{eq:zeta-choice}). Second, we will explore both options when comparing to electroproduction data. As we will see, the sign choice suggested by large-$N_c$ QCD leads to a better description of data. 

In the result sections we will also predict the radii that correspond to the TFFs. They are defined by \cite{Granados:2017cib}
\begin{equation}
    \langle r^2 \rangle_m := \frac6{G_m(0)} \left. \frac{\text{d} G_m(q^2)}{\text{d}q^2} \right\vert_{q^2=0}    \,.
    \label{eq:def-radii}
\end{equation}
In principle, this slope of a form factor at the photon point can be measured in the spacelike and in the timelike region. Concerning hyperons instead of Deltas and nucleons, it has been suggested in \cite{Granados:2017cib,Husek:2019wmt} to measure the radius in the timelike region using Dalitz decays. Valuable information about the structure of hyperons (TFFs in the spacelike region) can be extracted in this way. Here we can go the opposite way and predict the radius and the Dalitz decay distributions after exploring the spacelike region.

This brings us to the decay processes. The decay width of $\Delta \to n \gamma$ is given by
\begin{eqnarray}
  \Gamma_{\Delta \to n \gamma } & =& \frac{e^2 (m_{\Delta}^2-m_{n}^2)}{96 \pi m_{\Delta}^3} (m_{\Delta}-m_{n})^2
              \nonumber \\ 
  && \times \left( 3 |G_{+1}(0)|^2+|G_{-1}(0)|^2 \right)   \,.
  \label{eq:photCase}
\end{eqnarray}

We note in passing that the use of the constraint-free FFs, $F_i$, would generate  interference terms. Therefore, helicity amplitudes $G_m$ are much more convenient here. The Jones-Scadron FFs share the same feature on account of the relation
\begin{eqnarray}
&& 3 |G_{+1}(q^2)|^2+|G_{-1}(q^2)|^2  = \nonumber \\
&& \frac32 \left( \frac{m_\Delta + m_n}{m_n} \right)^2
\left( 3 |G^*_{E}(q^2)|^2+|G^*_{M}(q^2)|^2  \right)  \,.
    \label{eq:plmin-ME}
\end{eqnarray}

Next we provide the double differential decay rate for the Dalitz decay $\Delta \to n \; e^+ e^-$, first keeping the electron mass $m_e$ and then neglecting it:
\begin{eqnarray}
  &&\frac{\text{d}\Gamma_{\Delta \to n \; e^+ e^-}}{\text{d}q^2 \, \text{d}\cos\theta} =  \nonumber \\
  &&\frac{e^4}{(2\pi)^3 \, 96 m_{\Delta }^3 q^2} \, p_z \, \frac{\sqrt{q^2}}{2} \, \beta_e
  \, \left((m_{\Delta }-m_{n})^2-q^2\right)  \nonumber \\
  && \times \left[
     \left( 1+\cos^2 \theta + \frac{4m_e^2}{q^2}\, \sin^2 \theta \right) \right. \nonumber \\
  && \phantom{mn} \times \left( 3|G_{+1}(q^2)|^2+|G_{-1}(q^2)|^2 \right) 
  \nonumber \\
  && \phantom{m} \left. 
     {}+4\left(\sin^2 \theta + \frac{4m_e^2}{q^2}\cos^2 \theta \right)\frac{q^2}{m_{\Delta }^2} |G_0(q^2)| ^2\right]
       \nonumber \\[1em]
  && \approx \frac{e^4}{(2\pi)^3 \, 96 m_{\Delta }^3 q^2} \, p_z \, \frac{\sqrt{q^2}}{2} \, \beta_e
  \, \left((m_{\Delta }-m_{n})^2-q^2\right)  \nonumber \\
  && \phantom{m} \times \Big[
     \left( 1+\cos^2 \theta \right) \left( 3|G_{+1}(q^2)|^2+|G_{-1}(q^2)|^2 \right) 
  \nonumber \\
  && \phantom{mmn} 
     {}+\frac{4q^2}{m_{\Delta }^2}\sin^2 \theta \, |G_0(q^2)| ^2 \Big]  \,.
    \label{eq:Dalitzdistr}
\end{eqnarray}
The  kinematical velocity factor associated to the electron is defined as 
\begin{eqnarray}
  \label{eq:defbetaelectron}
  \beta_e := \sqrt{1-\frac{4m_e^2}{q^2}}. 
\end{eqnarray}
In Eq.\ \eqref{eq:Dalitzdistr}  the angle $\theta$ is taken between the electron and the neutron in the rest frame of the electron-positron pair. 

For later use we also introduce a QED version of \eqref{eq:Dalitzdistr}, which is supposed to describe the situation where the hyperon structure is not resolved; see also the discussion in \cite{Salone:2021bvx}. In practice we replace the TFF combinations by their $q^2=0$ expressions and make in this way also contact with the real photon case \eqref{eq:photCase}:
\begin{eqnarray}
  &&\frac{\text{d}\Gamma_{\Delta \to n \; e^+ e^-}^{\mathrm{QED}}}{\text{d}q^2 \, \text{d}\cos\theta} :=  \nonumber \\
  &&\frac{e^4}{(2\pi)^3 \, 96 m_{\Delta}^3 q^2} \, p_z \, \frac{\sqrt{q^2}}{2} \, \beta_e
  \, \left((m_{\Delta}-m_{n})^2-q^2\right)  \nonumber \\
  && \times 
     \left( 1+\cos^2 \theta + \frac{4m_e^2}{q^2}\, \sin^2 \theta \right)  \nonumber \\
  && \phantom{mn} \times \left( 3|G_{+1}(0)|^2+|G_{-1}(0)|^2 \right)\,.
    \label{eq:QEDcase}
\end{eqnarray}

Phenomenologically, it turns out that in the spacelike low-energy region, the TFFs of the electric and the Coulomb quadrupole are much smaller than the TFF of the magnetic dipole \cite{Pascalutsa:2006up}. This property is also true for the timelike low-energy region \cite{Ramalho:2015qna} (our results of Sections \ref{sec:results} and \ref{sec:outlook} support this statement). If one neglects electric and Coulomb TFFs (and the electron mass), then \eqref{eq:Dalitzdistr} can be approximated by 
\begin{eqnarray}
  &&\frac{\text{d}\Gamma_{\Delta \to n \; e^+ e^-}}{\text{d}q^2 \, \text{d}\cos\theta} \approx  \nonumber \\
  && \frac{e^4}{(2\pi)^3 \, 96 m_{\Delta }^3 q^2} \, p_z \, \frac{\sqrt{q^2}}{2} \, \beta_e
  \, \left((m_{\Delta }-m_{n})^2-q^2\right)  \nonumber \\
  && \phantom{m} \times 
     \left( 1+\cos^2 \theta \right) \frac32 \left( \frac{m_\Delta + m_n}{m_n} \right)^2 |G^*_{M}(q^2)|^2   \,.
    \label{eq:Dalitzdistr2}
\end{eqnarray}
In this approximation, the normalized angular distribution is trivial, i.e.\ independent of the (magnetic) TFF:
\begin{eqnarray}
  \frac{1}{\Gamma_{\Delta \to n \; e^+ e^-}} \, \frac{\text{d}\Gamma_{\Delta \to n \; e^+ e^-}}{\text{d}\cos\theta} \approx
  \frac38 \left( 1+\cos^2 \theta \right)  \,.
  \label{eq:triv-angle}
\end{eqnarray}
Thus all relevant information is contained in the singly-differential dilepton-mass distribution
$\text{d}\Gamma_{\Delta \to n \; e^+ e^-}/\text{d}q^2$, which we will predict in Subsection \ref{sec:res-Dalitz}. The (trivial) angular distribution is in  agreement with the findings of the HADES collaboration \cite{Adamczewski-Musch:2017hmp}.

\section{Dispersive machinery}
\label{sec:disp}

For completeness this section repeats   the general framework presented in \cite{Granados:2017cib,Junker:2019vvy}  but with particular attention towards the case studied here, the $\Delta$-$n$  transition. In order to incorporate the pion rescattering effect, we use the Omn\`es function, 
\begin{eqnarray}
  \Omega(s) = \exp\left\{ s \, \int\limits_{4m_\pi^2}^\infty \frac{\text{d}s'}{\pi} \, \frac{\delta(s')}{s' \, (s'-s-i \epsilon)} \right\}
  \label{eq:omnesele}  
\end{eqnarray}
where $\delta$ denotes the pion p-wave phase shift \cite{Colangelo:2001df,Garcia-Martin:2011iqs}. 

The pion vector FF, $F^V_\pi$, is taken 
from \cite{Leupold:2017ngs} (see also \cite{Hanhart:2012wi,Hanhart:2013vba,Hoferichter:2016duk,Alvarado:2023loi}):
\begin{eqnarray}
  \label{eq:FV-Omnes-alphaV}
  F^V_\pi(s) = (1+\alpha_V \, s) \, \Omega(s) \,.
\end{eqnarray}
For a value of $\alpha_V = 0.12 \, {\rm GeV}^{-2}$, equation \eqref{eq:FV-Omnes-alphaV} describes very well the pion vector FF data obtained from tau decays \cite{Belle:2008xpe} for energies below 1 GeV  \cite{Leupold:2017ngs}.

\subsection{Dispersion relations}
\label{subsec:dispgen}
As motivated in \cite{Junker:2019vvy}, we expect that the three TFFs $G_m$ introduced in \eqref{eq:F-1def}, \eqref{eq:F0def}, \eqref{eq:F+1def} and therefore also the Jones-Scadron TFFs of \eqref{eq:Gs-jones-scadron} satisfy unsubtracted dispersion relations:
\begin{eqnarray}
  \label{eq:disp-verygenGm}
  G_m(q^2) = \int \frac{\text{d}s}{2\pi i} \, \frac{{\rm disc}\,G_m(s)}{s-q^2}
\end{eqnarray}
for $m=0, \pm 1;M,E,C$. 

The low-energy behavior of the TFFs is determined by the lightest hadronic state that couples to the $\Delta\bar{n}$ system: the two-pion state.
Therefore, in complete analogy to \cite{Granados:2017cib}, we can write:
\begin{eqnarray}
  G_m(q^2) &=& \frac{1}{12\pi} \, \int\limits_{4 m_\pi^2}^\infty \frac{\text{d}s}{\pi} \, 
     \frac{T_m(s) \, p_{\rm c.m.}^3(s) \, F^{V*}_\pi(s)}{s^{1/2} \, (s-q^2-i \epsilon)}  \nonumber \\
     && {} + G^{\rm anom}_m(q^2) + \ldots
  \label{eq:genGmdispelips}
\end{eqnarray}
where the $\Delta \bar{n}\pi^+\pi^-$ scattering amplitudes are denoted by $T_m$. Here $p_{\rm c.m.}$ denotes the modulus of the momenta of the pions in the center-of-mass frame. The ellipsis stand for the infinitely many contributions coming from other intermediate states, such as four-pion \cite{Niecknig:2012sj}, two-kaon or baryon-antibaryon states. None of these will  be taken into account in this work; they are regarded as negligible as long as the TFFs are studied at low energies. To enhance further the importance of the low-energy region in the dispersive integral, one can introduce additional subtractions. We will come back to this aspect in Subsection \ref{subsec:sub}. 
The ``anomalous'' contribution $G^{\rm anom}_m$ will also be introduced later.

The pion-baryon scattering amplitudes $T_m$ are obtained in a two-step procedure. First 
we define the reduced amplitudes: 
\begin{eqnarray}
  K_{\pm 1}(s) & := & -\frac{3}{4} \, \int\limits_0^\pi \text{d}\theta \, \sin^2 \theta \, \nonumber \\ && {} \times 
  \frac{{\cal M}(s,\theta,1/2 \pm 1,1/2)}{\bar v_n(-p_z,1/2) \, \gamma_5 \, u^1_{\Delta}(p_z,1/2 \pm 1) \, p_{\rm c.m.} } \,,
  \nonumber \\
  K_0(s) & := & -\frac{3}{2} \, \frac{m_{\Delta}^2-m_n^2+s}{2 \, s} \, \int\limits_0^\pi \text{d}\theta \, 
  \sin \theta \, \cos\theta \nonumber \\ && {} \times 
  \frac{{\cal M}(s,\theta,1/2,1/2)}{\bar v_n(-p_z,+1/2) \, \gamma_5 \, u^3_{\Delta}(p_z,+1/2)\, p_{\rm c.m.} }  \,. \phantom{mm}
  \label{eq:def-redampl}
\end{eqnarray}
 We have introduced
${\cal M}(s,\theta,\sigma,\lambda)$ as the approximation to the Feynman amplitude for the reaction
$\Delta \, \bar n \to \pi^+ \pi^-$. In practice, ${\cal M}(s,\theta,\sigma,\lambda)$ does not include the rescattering
effect of the pions, but this will be taken care of in a second step. 
In addition, we want to distinguish conceptually between processes with
left-hand cut structures and purely polynomial terms \cite{Garcia-Martin:2010kyn,Kang:2013jaa}. Therefore, we will use the notation  $K_m$ to refer exclusively to the amplitudes that originate from the left-hand cut
structures (and drop at high energies), while we will denote the polynomial terms by $P_m$. In practice we will determine $K_m$ from the $\chi$PT tree-level expressions for nucleon and $\Delta$ exchange \cite{Granados:2017cib,Leupold:2017ngs,Junker:2019vvy,Alvarado:2023loi}. 

Pion rescattering is then taken into account by solving a Muskhelishvili-Omn\`es equation \cite{zbMATH03081975,Omnes:1958hv}.
The result is
\begin{eqnarray}
  T_m(s) & = & K_m(s) + \Omega(s) \, P_{m} + T_m^{\rm anom}(s) 
  \nonumber \\ && {}+ \Omega(s) \, s \, 
  \int\limits_{4m_\pi^2}^{\Lambda^2} \, \frac{\text{d}s'}{\pi} \, 
  \frac{K_m(s') \, \sin\delta(s')}{\vert\Omega(s')\vert \, (s'-s-i \epsilon) \, s'} \,. \phantom{m}  
  \label{eq:tmandel}
\end{eqnarray}
Note that we have introduced a cutoff $\Lambda$. Since we have only the low-energy part under control, where the
two-pion state dominates, it is not reasonable to extend the integral into the uncontrolled high-energy region. In practice,
the two-pion state dominates the isovector channel up to about 1 GeV. Beyond this point, four-pion states might also become important \cite{Niecknig:2012sj,Hanhart:2012wi}. To explore the uncertainties of our low-energy
approximation we will vary the cutoff between 1 and 2 GeV. 

We have used a once-subtracted dispersion relation in \eqref{eq:tmandel}. For the polynomial 
$P_m$ we just take a constant (per channel) that can be obtained from a fit to data. Note that in \eqref{eq:tmandel} this polynomial is multiplied
by the Omn\`es function $\Omega$. The latter drops at high energies. Therefore the restriction of $P_m$ to a constant has the additional feature
that $T_m$ drops for high energies. Certainly a benefit for the integrand of the dispersion relation \eqref{eq:genGmdispelips}.
Finally,  $T^{\rm anom}(s)$ denotes an additional contribution to the amplitude, associated to the presence of an anomalous cut on the first Riemann sheet.

In Appendix \ref{sec:anom-cutt-appendix} it is shown that the presence of an anomalous cut in the first Riemann sheet leads to a modification of the dispersion relations for both the amplitudes $T_m$ and the TFFs $G_m$. These additional terms are denoted by $T_m^{\text{anom}}$ and $G_m^{\text{anom}}$ and are provided in Appendix \ref{sec:anom-cutt-appendix}. As a consequence, the TFF integral in \eqref{eq:genGmdispelips} becomes complex for any $q^2$ value. Without, it would be real below the 
two-pion threshold.  Concretely, the anomalous pieces reflect the fact that the $\Delta$ is unstable, hence the exchanged  $p$-$\pi$ pair can be on-shell and therefore contribute to the imaginary part. This made it necessary to specify in \eqref{eq:GM-iMM} and \eqref{eq:multip-ratios} that the respective real part (and not, e.g., the modulus) enters the equations. For the decay formulae \eqref{eq:photCase} and \eqref{eq:Dalitzdistr}, on the other hand, it is the respective modulus that appears, see also the corresponding discussion in Appendix \ref{sec:exp-re-BW}. 

Some comments about other inelasticities are in order. 
As motivated already in \cite{Junker:2019vvy},  we do not include the kaons as intermediate states. The two-kaon threshold starts at  $(2 m_{K})^2 \approx 1$ GeV$^2$, rather far away from the two-pion threshold which is located  at $(2 m_\pi)^2 \approx 0.08$ GeV$^2$. The latter is the most relevant from a dispersive point of view, since the influence of high-energy inelasticities is naturally suppressed for low values of $q^2$.
The branch point of the anomalous cut associated to the proton-pion-pion triangle lies in the vicinity, i.e.\ at $s_+ \approx (0.05 -0.08 i)$ GeV$^2$. It is therefore also important. Finally, note that there cannot be additional  anomalous cuts since the reaction $\Delta \bar n \to K \bar K$ requires the exchange of hyperons, which are too heavy to satisfy the condition \eqref{eq:anomthr}. 

A second reason for not including the two-kaon state lies in the fact that the four-pion state seems to be more important than the two-kaon state \cite{Hanhart:2012wi,Niecknig:2012sj} (and both seem to be of minor importance at sufficiently low energies). The two-kaon inelasticity (but not the four-pion states) has been included in a recent analysis of the TFFs of $\Sigma^0$ to $\Lambda$ \cite{Lin:2022dyu}. It is encouraging that it has been confirmed in \cite{Lin:2022dyu} that the influence of the two-kaon states is minor.

\subsection{Subtracted dispersion relations}
\label{subsec:sub}

A once-subtracted dispersion relation is used to enhance
the importance of the low-energy region in the dispersive integral: 
\begin{eqnarray}
  &&  G_m(q^2) =  G_m(0)  
     \nonumber \\
  && {}
     +  \frac{q^2}{12\pi} \, \int\limits_{4 m_\pi^2}^{\Lambda^2} \frac{\text{d}s}{\pi} \, 
     \frac{T_m(s) \, p_{\rm c.m.}^3(s) \, F^{V*}_\pi(s)}{s^{3/2} \, (s-q^2-i \epsilon)}
     + G^{\rm anom}_m(q^2) 
     \label{eq:dispbasic}  
\end{eqnarray}
for $m=0,\pm 1, M,E,C$. In principle, the three subtraction constants $G_m(0)$ are complex-valued, but at least the real parts can be determined by experiment. This is a great advantage compared to the $\Sigma^*$-$\Lambda$ case \cite{Junker:2019vvy}, where only unsubtracted dispersion relations could be used due to a lack of data. The subtraction constants encode the high-energy contributions left out from our formalism; see also the corresponding discussions in \cite{Granados:2017cib,Hoferichter:2016duk,Leupold:2017ngs,Junker:2019vvy,Alvarado:2023loi}.  
The last,  ``anomalous'' piece in \eqref{eq:dispbasic} is given by
\begin{eqnarray}
  G^{\rm anom}_m(q^2) &=& \frac{q^2}{12\pi} \, \int\limits_{0}^1 \text{d}x \, \frac{ds''(x)}{dx} \,
                          \frac{1}{s''(x)-q^2}   \nonumber \\ && \times
  \frac{f_m(s''(x)) \, F^{V}_\pi(s''(x))}{-4\, (-\lambda(s''(x),m_{\Delta}^2,m_n^2))^{3/2}}  \,.
  \label{eq:Fanom}
\end{eqnarray}
More technical details on this quantity are provided in Appendix \ref{sec:anom-cutt-appendix}. 

Finally, we come back to the kinematical constraints \eqref{eq:kinconstr00}, \eqref{eq:constr-EC*} that helicity amplitudes and Jones-Scadron TFFs must satisfy. If we had a full coverage of all inelasticities (and not only two pions) in the dispersion relation \eqref{eq:genGmdispelips} and if we knew all scattering amplitudes up to very large energies, then the kinematical constraints would be automatically satisfied. In practice, we have only control over the low-energy region with its dominance of the two-pion state. Therefore, the kinematical constraints will be violated to some extent. There are several ways how to deal with this issue. If one formulates dispersion relations for the constraint-free TFFs, $F_{1,2,3}$, introduced in \eqref{eq:defTFF2}, then the helicity amplitudes will be obtained from \eqref{eq:F-1def}, \eqref{eq:F0def}, and \eqref{eq:F+1def}. The kinematical constraints will then be automatically satisfied. This is the path followed in \cite{Alvarado:2023loi,AnDi-prep}. The disadvantage for the system studied here is the fact that it is then difficult to disentangle small and large TFFs. Technically it is also simpler to use helicity amplitudes for the hadronic input. 
A second option is the use of subtracted dispersion relations where the subtraction constants are subject to the kinematical constraints. Of course, one loses some freedom to parametrize the unknown high-energy physics.

In any case, the kinematical constraint \eqref{eq:constr-EC*} does not touch the dominant magnetic TFF but concerns solely the numerically much smaller TFFs $G_E^*$ and $G_C^*$. We leave it to future work to explore in detail which dispersion relations are best suited for these electric and Coulomb quadrupole TFFs.

\section{Input from chiral perturbation theory}
\label{sec:chiPT}

\subsection{Effective Lagrangians}
\label{sec:eff-lagr}

The leading-order (LO) chiral Lagrangian 
including the decuplet states 
is given by \cite{Jenkins:1991es,Lutz:2001yb,Pascalutsa:2006up,Ledwig:2014rfa,Holmberg:2018dtv,Mommers:2022dgw} 
\begin{eqnarray}
  && {\cal L}_{\rm baryon}^{(1)} = {\rm tr}\left(\bar B \, (i \slashed{D} - m_{(8)}) \, B \right)  \nonumber \\ 
  && {}+ \bar T_{abc}^\mu \, ( i \gamma_{\mu\nu\alpha} D^\alpha - \gamma_{\mu\nu} \, m_{(10)}) \, (T^\nu)^{abc}
  \nonumber \\ 
  && {}+ \frac{D}{2} \, {\rm tr}(\bar B \, \gamma^\mu \, \gamma_5 \, \{u_\mu,B\}) 
  + \frac{F}{2} \, {\rm tr}(\bar B \, \gamma^\mu \, \gamma_5 \, [u_\mu,B])  \nonumber \\
  && {} + \frac{h_A}{2\sqrt{2}} \, 
  \left(\epsilon^{ade} \, \bar T^\mu_{abc} \, (u_\mu)^b_d \, B^c_e
  + \epsilon_{ade} \, \bar B^e_c \, (u^\mu)^d_b \, T_\mu^{abc} \right) \nonumber \\
  && {} - \frac{H_A}{4 m_R} \, \epsilon_{\mu\nu\alpha\beta} \, 
  \left(\bar T^\mu_{abc} \, (D^\nu T^\alpha)^{abd} \, (u^\beta)^c_d \right. \nonumber \\
  && \left. \phantom{mmmmmmmm} + (D^\nu \bar T^\alpha)_{abd} \, (T^\mu)^{abc} \, (u^\beta)^d_c \right)  \,.
  \label{eq:baryonlagr}
\end{eqnarray}
Here $B$ contains the states of the baryon octet, $T$ collects the baryon decuplet states and $u_\mu$ contains the Goldstone boson fields. For further details we refer to \cite{Junker:2019vvy,Holmberg:2018dtv,Mommers:2022dgw}.

In (\ref{eq:baryonlagr}) $m_{(8)}$ ($m_{(10)}$) denotes the mass of the baryon octet (decuplet) in the chiral limit. 
For the next-to-leading-order (NLO) calculation that we perform in the present work we use the 
physical masses \cite{ParticleDataGroup:2022pth} of all states. However, since our accuracy is not high enough to resolve isospin breaking effects, we  take one average mass for proton and neutron,
e.g.\ $m_p \approx m_n \approx m_N=0.939 \,$GeV. We use for the $\Delta$-resonance mass $m_\Delta = 1.232\,$GeV \cite{ParticleDataGroup:2022pth} and for the pion mass  $m_\pi=0.13957\,$GeV.

In line with \cite{Junker:2019vvy} we use $F_\pi = 92.28\,$MeV and  $D=0.80$, $F=0.46$, which implies for the pion-nucleon coupling constant $g_A=F+D =1.26$. Sizes and signs of $D$ and $F$ have been obtained from the weak semileptonic decays of octet baryons. 

For the three-point coupling constants $h_A$ and $H_A$ we use the two-flavor estimates for a 
large number of colors \cite{Pascalutsa:2005nd}: 
$\vert h_A \vert \approx 3g_A/\sqrt{2} \approx 2.67$ and $H_A \approx 9g_A/5 \approx 2.27$. 
This choice is slightly different from \cite{Junker:2019vvy} where we used $h_A$ as determined from hyperon decays, cf.\ also the 
discussion in \cite{Holmberg:2018dtv,Holmberg:2019ltw}. 
In the present paper we explore solely the hadron sector of up and down quarks. Therefore two-flavor estimates are more 
reasonable. The size of $h_A$ fits rather well to its determination from the decay width of $\Delta \to N \pi$ \cite{Granados:2017cib,Leupold:2017ngs,Holmberg:2018dtv,Holmberg:2019ltw,Junker:2019vvy}.

As already pointed out, one has a free phase choice (sign choice) when introducing a decuplet field. One could use this freedom to choose a positive or negative value for $h_A$. Instead we have decided to use the freedom to adjust our TFFs to the experimental convention. This is the essence of the choice (\ref{eq:zeta-choice}). 
Therefore we have to explore which sign our coupling constant $h_A$ should have. As often in theoretical physics, the choices are: calculate the sign of $h_A$ or fit it to data. We will explore both options. As a calculational tool we have decided to use QCD for a large number of colors. This calculation is provided in Appendix \ref{sec:largeNc}. When comparing to data in Section \ref{sec:results} we will see that the fit agrees with the sign prediction of large-$N_c$ QCD. 

We take this as an encouragement to use even the numerical prediction for the other three-point coupling, $H_A$. There, the sign is not a convention. It is correlated to the sign of $g_A$; see also the discussion in \cite{Bertilsson:2023htb}. This is easy to see when anticipating the calculations that we will present below. The reaction $\Delta \pi \to N \pi$ can proceed via nucleon exchange, where the amplitude is $\sim h_A g_A$, and also via $\Delta$ exchange, where the amplitude is $\sim H_A h_A$. These two amplitudes interfere. Thus the sign of $H_A$ relative to the sign of $g_A$ is not a matter of convention. On the other hand, the sign of $g_A$ is determined from the ``$v$ minus $a$'' structure of the weak interaction (and a sign convention for the vector part).  

Now we turn to the Lagrangian of second order in the chiral counting.
A complete and minimal NLO Lagrangian has been presented in \cite{Holmberg:2018dtv}. For our present purpose we need terms
that lift the mass degeneracies that hold at LO and we need terms that provide interactions for 
$\Delta \pi \to n \pi$ (or formally $\Delta \bar{n} \to 2\pi$) with the two pions in a p-wave. 

The relevant part of the NLO Lagrangian for the baryon octet sector reads \cite{Oller:2006yh,Frink:2006hx,Holmberg:2018dtv}
\begin{eqnarray}
  {\cal L}_{8}^{(2)} &=& b_{\chi,D} \, {\rm tr}(\bar B \, \{\chi_+,B\}) + b_{\chi,F} \, {\rm tr}(\bar B \, [\chi_+,B]) 
  \label{eq:octetNLO}
\end{eqnarray}
with $\chi_\pm = u^\dagger \chi u^\dagger \pm u \chi^\dagger u$ and $\chi = 2 B_0 \, (s+ip)$ 
obtained from the scalar source $s$ and the pseudoscalar source $p$. The low-energy constant $B_0$ is essentially the ratio of
the light-quark condensate and the square of the pion-decay constant; 
see, e.g.\ \cite{Gasser:1983yg,Gasser:1984gg,Scherer:2002tk,Scherer:2012xha}. 
While at LO all baryon octet states are degenerate in mass, the NLO terms of (\ref{eq:octetNLO}) lift this degeneracy and 
essentially move all masses to their respective physical values. Technically this is achieved if one replaces the 
scalar source $s$ by the quark mass matrix. Numerical results for the octet mass $m_{(8)}$ 
in (\ref{eq:baryonlagr}) and the splitting parameters $b_{\chi,D/\chi,F}$ in (\ref{eq:octetNLO}) 
are given, for instance, in \cite{Kubis:2000aa}. In practice we use the physical masses. Therefore we do not specify these 
parameters here.

The relevant part of the NLO Lagrangian for the baryon decuplet sector reads \cite{Holmberg:2018dtv}
\begin{eqnarray}
  {\cal L}_{10}^{(2)} &=& -d_{\chi,(8)} \bar T_{abc}^\mu \, (\chi_+)^c_d \, \gamma_{\mu\nu} \, (T^\nu)^{abd}  \,.
  \label{eq:decupletNLO}
\end{eqnarray}
It provides a mass splitting for the decuplet baryons such that 
$m_\Omega - m_{\Xi^*} = m_{\Xi^*} - m_{\Sigma^*} =  m_{\Sigma^*} - m_\Delta$, in good agreement with phenomenology \cite{ParticleDataGroup:2022pth}. 
In the present work we only deal with the $\Delta$ and, in practice, we use the physical mass of the neutral $\Delta$. In that way the physical thresholds are exactly reproduced. 

For the formal reaction $\Delta^0 \bar{n} \to \pi^+ \pi^-$ the relevant part of the NLO Lagrangian \cite{Holmberg:2018dtv} 
is given by
\begin{eqnarray}
  \label{eq:NLOtrans}
  {\cal L}_{8-10}^{(2)} & \to & -\frac{c_F}{\sqrt{3}F_\pi^2} \, \bar n \gamma_\mu \gamma_5 \Delta_\nu^{0} 
  \left(\partial^\mu \pi^+\,\partial^\nu \pi^- - (\mu \leftrightarrow \nu) \right)   \,. 
\end{eqnarray}
One could fit $c_F$ to data when comparing calculation and experimental results for ${\rm Re}G^*_M$ as given in (\ref{eq:GM-iMM}). The low-energy constant $c_F$ will contribute to $P^*_M$, see Appendix \ref{sec:contact-ChPT}. On the other hand, one can fit the constants $P_m$ right away to data. 

To judge the quality of our reproduction of data and of our predictions, it is important to understand how well our theory parameters
are constrained and to check how much the observables are sensitive to parameter variations. Therefore, a brief summary of our theory
parameters is in order. We use isospin averaged (but not three-flavor averaged) hadron masses throughout. The pion decay constant $F_\pi$ and the axial charge of the nucleon $g_A$ are very well known from weak decays. We will not explore parameter variations of masses or weak-decay constants.

The subtraction constants $G_m(0)$ and $P_m$ appearing in \eqref{eq:dispbasic} and \eqref{eq:tmandel}, respectively, will be fitted directly to spacelike TFF data. We will check, however, which value for $G_M^*(0)$ will be provided by an unsubtracted dispersion relation.

The remaining input parameters that are not fitted to TFF data are the $\Delta$-$N$-$\pi$ coupling constant $\vert h_A \vert$, the (p-wave) $\Delta$-$\Delta$-$\pi$ coupling constant $H_A$, and the cutoff $\Lambda$ that appears in the dispersive integrals \eqref{eq:tmandel} and \eqref{eq:dispbasic}. As already spelled out, we will vary $\Lambda$ between 1 and $2\,$GeV. For the three-point coupling constants we will explore variations of about 10\% up and down, i.e.
\begin{eqnarray}
  \label{eq:param-var-hA-HA}
  && 1 \le \Lambda \, [{\rm GeV}] \le 2 \,, \nonumber \\
  && 2.4 \le \vert h_A \vert \le 2.9 \,, \nonumber \\
  && 2.0 \le H_A \le 2.5  \,.
\end{eqnarray}

\subsection{Matrix elements}

The $\Delta\bar{n}\pi^+\pi^-$ tree-level amplitudes, i.e.\ $\chi$PT amplitudes up to (including) NLO, are calculated in a first step. 
Then, the reduced amplitudes are obtained applying the projection method presented 
in the appendix of Ref.\ \cite{Junker:2019vvy}.  The calculations are performed using FeynCalc \cite{Mathematica, MERTIG1991345}.  Note that these amplitudes constitute the leading i.e.\ dominant contribution; the notation NLO refers to the underlying chiral Lagrangian, whose tree-level contribution is equally important to that coming from the LO Lagrangian. Further comments on the power counting can be found in \cite{Junker:2019vvy}.

\begin{widetext}
The Feynman matrix element for the reaction $\Delta^{0} \bar n \to \pi^+ \pi^-$ up to (including) NLO is given by
\begin{eqnarray}
  && - \frac{g_A h_A}{2 \sqrt{6} F_\pi^2} \,  \frac{1}{u-m_p^2+i\epsilon} \, 
  p^\mu_{\pi^-} g_{\mu\alpha} \, \bar v_n \, \slashed{p}_{\pi^+} \gamma_5 \, 
  (\slashed{p}_{\Delta} - \slashed{p}_{\pi^-} + m_p) \, u^\alpha_{\Delta}
  \nonumber \\
   && {} + \frac{h_A H_A}{3 \sqrt{6} m_{\Delta} F_\pi^2} \, i \epsilon^\lambda_{\phantom{\lambda}\nu\alpha\beta} \, 
  p^\nu_{\Delta} \, p^\beta_{\pi^-} \, p^\mu_{\pi^+} \, \bar v_n S_{\mu\lambda}(p_{\Delta}-p_{\pi^-}) \, u^\alpha_{\Delta} 
  \nonumber \\
  && {}- \frac{h_A H_A}{2 \sqrt{6} m_{\Delta} F_\pi^2} \, i \epsilon^\lambda_{\phantom{\lambda}\nu\alpha\beta} \, 
  p^\nu_{\Delta} \, p^\beta_{\pi^+} \, p^\mu_{\pi^-} \, \bar v_n S_{\mu\lambda}(p_{\Delta}-p_{\pi^+}) \, u^\alpha_{\Delta} 
  \nonumber \\
  && {} - \frac{c_F}{\sqrt{3} F_\pi^2} \, (p^\mu_{\pi^+} p^\alpha_{\pi^-} - p^\alpha_{\pi^+} p^\mu_{\pi^-} ) \, g_{\alpha\beta} \, 
  \bar v_n \gamma_\mu \gamma_5 u^\beta_{\Delta}   \,.
  \label{eq:feynlo+nlo}
\end{eqnarray}
Here $S_{\mu\nu}$ denotes the spin-3/2 propagator  \cite{deJong:1992wm,Hacker:2005fh}:
\begin{eqnarray}
  \label{eq:defspin32prop2}
  S_{\mu\nu}(p) & := & - \frac{\slashed{p}+m_\Delta}{p^2-m_\Delta^2+ i\epsilon} \, P^{3/2}_{\mu\nu}(p) 
  + \frac{2}{3m_\Delta^2} \, (\slashed{p}+m_\Delta) \, \frac{p_\mu p_\nu}{p^2}   \nonumber \\ 
  && {} - \frac{1}{3m_\Delta} \, \frac{p_\mu \, p^\alpha \, \gamma_{\alpha\nu} + \gamma_{\mu\alpha} \, p^\alpha \, p_\nu}{p^2}  \,. 
\end{eqnarray}

The reduced amplitudes associated to the $\Delta^\pm$ and proton exchange diagrams constitute the bare $\chi$PT input. They are given below in the form $K_m+P_m$, where $P_m$ are constant terms that are left out of the dispersive integrals. In line with \cite{Leupold:2017ngs} we have also found here that it is most reasonable to determine $P_m$ by fits to data. But for completeness we provide the $\chi$PT expressions for $P_m$ in Appendix \ref{sec:contact-ChPT}. 

The explicit expressions for the left-hand-cut contributions are
\begin{eqnarray}
  K_{+1} & = & -\frac{g_A h_{A}}{4 \sqrt{6} F_{\pi }^2}( C_{+1} + D_{+1} \, R^{\rm oct.}_s)
  - \frac{5h_{A} H_{A}}{12 \sqrt{6} F_{\pi }^2} \, ( E_{+1} + F_{+1} \, R^{\rm dec.}_s) \,, \nonumber \\
  K_{-1} & = &- \frac{g_A h_{A}}{4 \sqrt{6} F_{\pi }^2}( C_{-1} + D_{-1} \, R^{\rm oct.}_s)
  - \frac{5h_{A} H_{A}}{12 \sqrt{6} F_{\pi }^2} \, ( E_{-1} + F_{-1} \, R^{\rm dec.}_s)  \,, \nonumber \\
  K_{0} & = &- \frac{g_A h_{A}}{4 \sqrt{6} F_{\pi }^2}( C_{0} + D_{0} \, R^{\rm oct.}_d)
  - \frac{5h_{A} H_{A}}{12 \sqrt{6} F_{\pi }^2} \, ( E_{0} + F_{0} \, R^{\rm dec.}_d) 
  \label{eq:K-pw}  
\end{eqnarray}
with 
\begin{eqnarray}
  R^{\rm oct.}_s & = &
  \frac{-2Y_{p}}{\kappa^2} \left(1-\left(1-\frac{Y_{p}^2}{\kappa^2}\right) \frac{|\kappa|}{Y_{p}} 
    \left(\arctan\left(\frac{|\kappa|}{Y_{p}}\right)+\pi  \Theta (s_Y-s)\right)\right) \,, \nonumber \\[0.8em]
  R^{\rm oct.}_d & = & \frac{4}{\kappa^2} \left(1-\frac{Y_{p}}{|\kappa|} 
    \left(\arctan\left(\frac{|\kappa|}{Y_{p}}\right)+\pi  \Theta (s_Y-s)\right)\right)  \,, \nonumber \\[0.8em]
  R^{\rm dec.}_s & = & \frac{-2Y_{\Delta}}{\kappa^2} \left(1-\left(1-\frac{Y_{\Delta}^2}{\kappa^2}\right) 
    \frac{|\kappa|}{Y_{\Delta}} \, \arctan\left(\frac{|\kappa|}{Y_{\Delta}}\right) \right) \,, \nonumber  \\[0.8em]
  R^{\rm dec.}_d & = & \frac{4}{\kappa^2} \left(1-\frac{Y_{\Delta}}{|\kappa|} 
    \, \arctan\left(\frac{|\kappa|}{Y_{\Delta}}\right)\right)
  \label{eq:defRsd}
\end{eqnarray}
\end{widetext}
and
\begin{eqnarray}
  \label{eq:defABs}
  Y_{p} & = & 2 m_p^2-m_{\Delta}^2-m_n^2-2 m_\pi^2+s \,, \\
  Y_{\Delta} & = &  m_{\Delta}^2-m_n^2-2 m_\pi^2+s \,,  \\
  \kappa^2 & = & \frac{1}{s} \, (s-4 m_\pi^2) \, \lambda(s,m_{\Delta}^2,m_n^2)   \,, \\
  s_Y & = & m_{\Delta}^2+m_n^2+2 m_\pi^2-2 m_p^2  \,.
\end{eqnarray}
Further details are provided in Appendix \ref{sec:anom-cutt-appendix} and in \cite{Junker:2019vvy}.
Note that $\kappa^2$ is negative in the range $s_{dt} < s < s_{st}$, i.e.\ $\vert \kappa \vert = \sqrt{-\kappa^2}$.
Only for negative $\kappa^2$ the expressions \eqref{eq:defRsd} are correct. For positive $\kappa^2$ one has log's instead of arctan's. In practice, however, the integration boundaries of the dispersive integrals run from $4m_\pi^2$ to $\Lambda^2=4$ GeV$^2$ which lies below the scattering threshold $s_{st}=(m_{\Delta} + m_n)^2$. It follows that one has to replace the arctan's with the log's only in the range $4m_\pi^2<s<s_{dt}=(m_{\Delta} -m_n)^2$.

The coefficient functions in \eqref{eq:K-pw} are given by
\begin{align}
  C_{+1} = {} & - \frac{2 \, (m_{\Delta} - m_n) \, (m_n + m_p)}{s-\left(m_{\Delta}-m_n\right){}^2}  \,,
  \\
  C_{-1} = {} & - \frac{6\, (m_{\Delta} - m_n) \, (m_n + m_p)}{s-\left(m_{\Delta}-m_n\right){}^2}
  \,, \\
  C_{0} = {} & \frac{(m_{\Delta}+m_n) \, (m_{\Delta}+m_p)}{s} 
  -\frac{3 m_{\Delta} (m_n + m_p)}{s-(m_{\Delta}-m_n)^2}
\,,
\end{align}
\begin{widetext}
\begin{align}
  D_{+1} = {} & 3 m_p \, (m_n+m_p) + \frac{3 \, (m_{\Delta}-m_n) \, (m_n+m_p) \, 
  (m_\pi^2 + m_{\Delta} m_n - m_p^2)}{s-(m_{\Delta}-m_n)^2}
  \,, \\
  D_{-1} = {} & \frac{3}{m_{\Delta}} \, (m_n+m_p) \, (m_\pi^2 -m_{\Delta}^2 + m_{\Delta} m_p - m_p^2)
  \nonumber \\[0.7em] & {}
                + \frac{9 \, (m_{\Delta}-m_n) \, (m_n+m_p) \, 
  (m_\pi^2 + m_{\Delta} m_n - m_p^2)}{s-(m_{\Delta}-m_n)^2}
  \,, \\
  D_{0} = {} & 3m_p(m_n+m_p)(m_{\Delta}^2-m_{\Delta}m_p-m_\pi^2+m_p^2)
               -\frac{9m_{\Delta}(m_n+m_p)(m_{\Delta}m_n+m_\pi^2-m_p^2)^2}{s-(m_{\Delta}-m_n)^2} \nonumber \\
  & +\frac{3(m_{\Delta}+m_n)(m_p+m_n)}{s}\Big(m_{\Delta}^3m_n-m_p(m_{\Delta}-m_n)(m_{\Delta}^2+m_\pi^2)+2m_{\Delta}^2m_\pi^2 \nonumber \\
              & \phantom{mm}
 -m_p^2\big(m_{\Delta}(m_{\Delta}+m_n)+2m_\pi^2\big)+2m_{\Delta}m_n m_\pi^2-m_p^3(m_n-m_{\Delta})+m_\pi^4+m_p^4\Big)  \,,
\end{align}
\begin{align}
  E_{+1} = {} & \frac{(m_{\Delta}-m_n) \left((m_{\Delta}+m_n)^2- m_\pi^2\right)}%
  {3 m_{\Delta} \left(s-\left(m_{\Delta}-m_n\right){}^2\right)}
  \,, \\
  E_{-1} = {} & \frac{(m_{\Delta}-m_n) \left((m_{\Delta}+m_n)^2-m_\pi^2 \right)}{m_{\Delta} (s-(m_{\Delta}-m_n)^2)} \,, \\
  E_{0} = {} & - \frac{( m_{\Delta}+m_n) (2m_{\Delta}^2+2m_{\Delta}m_n -m_\pi^2)}{6 m_{\Delta}\, s}
 + \frac{(m_{\Delta}+m_n)^2-m_\pi^2}
 {2 (s-(m_{\Delta}-m_n )^2 )}  \,,
\end{align}
\begin{align}
  F_{+1} = {} & -\frac{3s}{2}-\frac{m_\pi^2(2m_{\Delta}+3m_n)}{2m_{\Delta}}+\frac{5(m_{\Delta}+m_n)^2}{2}
   \nonumber \\ &
  {}+\frac{(m_{\Delta}-m_n)((m_{\Delta}+m_n)^2-m_\pi^2)(m_{\Delta}^2-m_{\Delta}m_n-m_\pi^2)}{2m_{\Delta}(s-(m_{\Delta}-m_n)^2)}  \,, \\
  F_{-1} = {} & \frac{3s}{2} +\frac{m_\pi^2( m_{\Delta}^2+ m_{\Delta} m_n-m_n^2)+m_\pi^4}{2 m_{\Delta}^2} -\frac{5( m_{\Delta}+m_n)^2}{2} \nonumber \\
  & +\frac{3(m_{\Delta}-m_n)((m_{\Delta}+m_n)^2-m_\pi^2)(m_{\Delta}^2-m_{\Delta}m_n-m_\pi^2)}{2m_{\Delta}(s-(m_{\Delta}-m_n)^2)}   \,, \\
  F_{0} = {} & \frac{3m_{\Delta}^2\, s}{2} -\frac{m_\pi^2(7m_{\Delta}^2-2m_{\Delta}m_n+2m_n^2)+m_{\Delta}^2(m_{\Delta}+m_n)^2}{2}+m_\pi^4  \nonumber \\
  & +\frac{4m_{\Delta}^2m_\pi^2(m_{\Delta}-2m_n)(m_{\Delta}+m_n)^2-m_\pi^4(2m_{\Delta}^3+m_{\Delta}^2m_n+m_n^3)+m_\pi^6(m_{\Delta}+m_n)}{2m_{\Delta} \, s} \nonumber \\
  & +\frac{3((m_{\Delta}+m_n)^2-m_\pi^2)(m_{\Delta}(m_n-m_{\Delta})+m_\pi^2)^2}{2(s-(m_{\Delta}-m_n)^2)}   \,.
\end{align}
\end{widetext}

\section{Results}
\label{sec:results}

\subsection{Magnetic dipole transition form factor}
\label{sec:results-magTFF}

\begin{figure}
    \centering
     \includegraphics[width=0.6\textwidth]{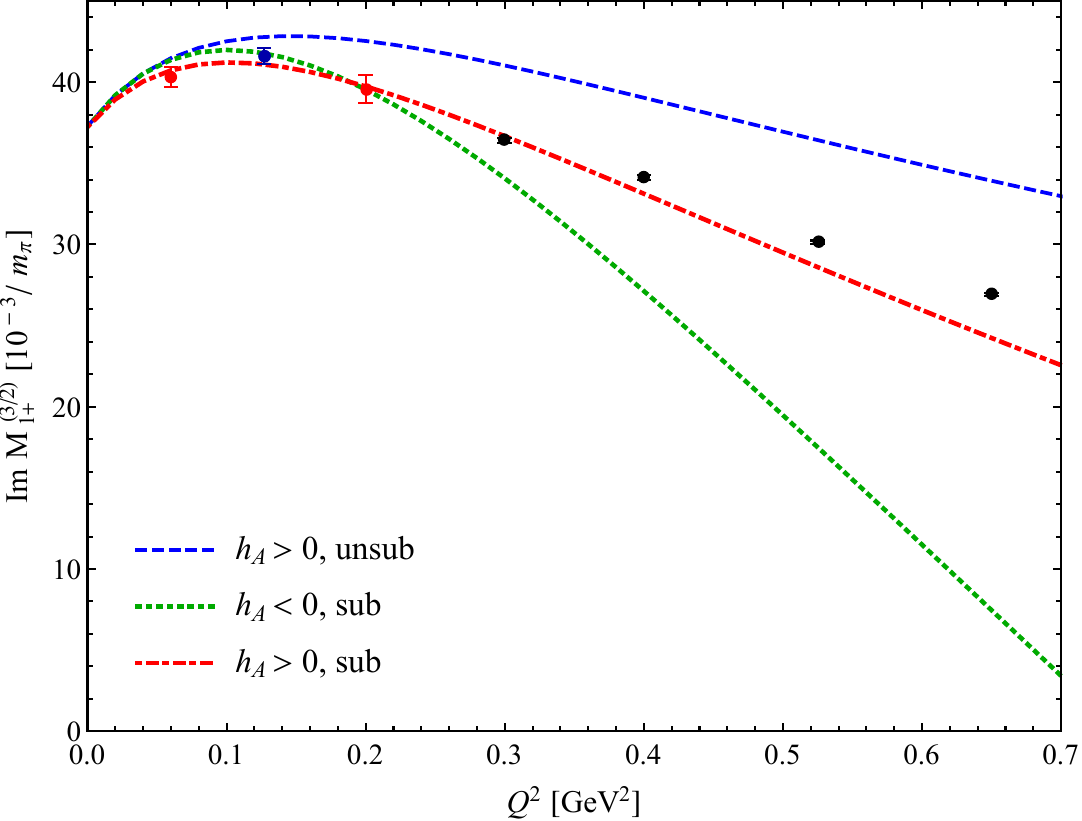}
    \caption{${\rm Im}M_{1+}^{(3/2)}$ as a function of $Q^2=-q^2$ using an unsubtracted dispersion relation with $h_A > 0$ (blue, dashed line),
    a subtracted dispersion relation with $h_A < 0$ (green, dotted), and 
    a subtracted dispersion relation with $h_A > 0$ (red, dash-dotted). Data taken from BATES (blue) \cite{OOPS:2004kai}, MAMI (red) \cite{Sparveris:2006uk,Stave:2006ea}, and CLAS (black) \cite{CLAS:2009ces}. The values of the contact terms are $P_{M}^* \approx 367\,$GeV$^{-2}$ (blue), $P_{M}^* \approx 570\,$GeV$^{-2}$ (green), and $P_{M}^* \approx 406\,$GeV$^{-2}$ (red), respectively.
}
    	\label{fig:ImM-various-signs}
\end{figure}

In Fig.\ \ref{fig:ImM-various-signs} we show the results for the quantity ${\rm Im}M_{1+}^{(3/2)}$ using a subtracted or unsubtracted dispersion relation, respectively, and varying the sign of $h_A$. For all cases, $P_M^*$ is used as a fit parameter. We observe that the unsubtracted dispersion relation provides only a qualitative description of the data. Quantitatively reasonable results are obtained for a subtracted dispersion relation. We also see that the sign choice $h_A > 0$ fits the data much better, in line with the large-$N_c$ considerations of Appendix \ref{sec:largeNc}. In the following, we will use only $h_A >0$. For the best fit, i.e.\ for the subtracted dispersion relation using $h_A > 0$ we find $P_M^* = 406.12\,\text{GeV}^{-2}$. The overall size of this subtraction constant is completely natural. We can deduce from the formulae of Appendix \ref{sec:contact-ChPT} that the scale is set by $1/F_\pi^2$ which is about $10^2 \,\text{GeV}^{-2}$. 

The agreement between theory and experiment reaches up to about $\sqrt{Q^2} \approx 0.6\,$GeV. This is achieved by only two fit parameters, the subtraction constant $P_M^*$ for the hadronic amplitude and the subtraction constant $G_M^*(0)$. Strictly speaking, the latter is just fitted to the experimental value at the photon point \cite{ParticleDataGroup:2022pth,Pascalutsa:2006up}, not to the spacelike data of Fig. \ref{fig:ImM-various-signs}. If we focus for a moment on the values around the photon point, we have to state that it is non-trivial to obtain a good reproduction of function value, slope and curvature(!) with only two fit parameters. Indeed, at low energies, the whole shape of the curve suggested by the experimental points is reproduced very well.  

\begin{figure}
    \centering
    \includegraphics[width=0.6\textwidth]{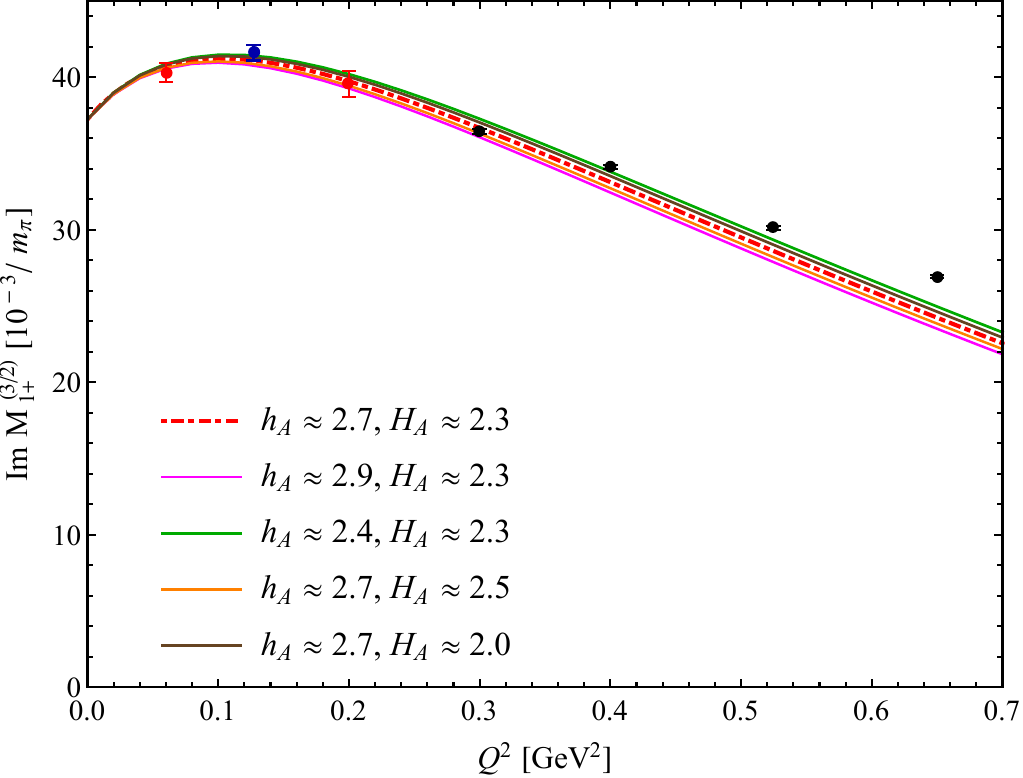}
    \caption{Same as Fig.\ \ref{fig:ImM-various-signs} for a subtracted dispersion relation and $h_A > 0$ but varying the input parameters. From top to bottom:
    $h_A$ decreased by 10\%, central value for $H_A$ (green); 
    $H_A$ decreased by 10\%, central value for $h_A$ (brown);
    central values for $h_A$ and $H_A$ (red, dash-dotted); 
    $H_A$ increased by 10\%, central value for $h_A$ (orange); 
    $h_A$ increased by 10\%, central value for $H_A$ (magenta).}
    	\label{fig:hAvariation}
\end{figure}

In Fig.\ \ref{fig:hAvariation} we show the impact of parameter variations. In general, the results are very stable if one changes one of the three-point coupling constants by 10\%. It is very encouraging that our results are so robust. We do not study variations of the cutoff $\Lambda$ in the plots to keep the number of lines manageable. Below we will explore such variations when providing values for the various radii. For Figs.\ \ref{fig:ImM-various-signs} and \ref{fig:hAvariation} we use $\Lambda = 2\,$GeV. Quantitatively, we see that an increase (decrease) in $h_A$ or $H_A$ leads to a decrease (increase) for ${\rm Im}M_{1+}^{(3/2)}$ in this range of negative $q^2$ (i.e.\ positive $Q^2$). This means that all three contributions (triangle diagrams with nucleon and with $\Delta$ and the contact term $P_M^*$) add up constructively to the slope of $G_M^*(q^2)$. We will see this more directly in Table \ref{table:radii-res} below. 

\begin{figure}
    \centering
     \includegraphics[width=0.6\textwidth]{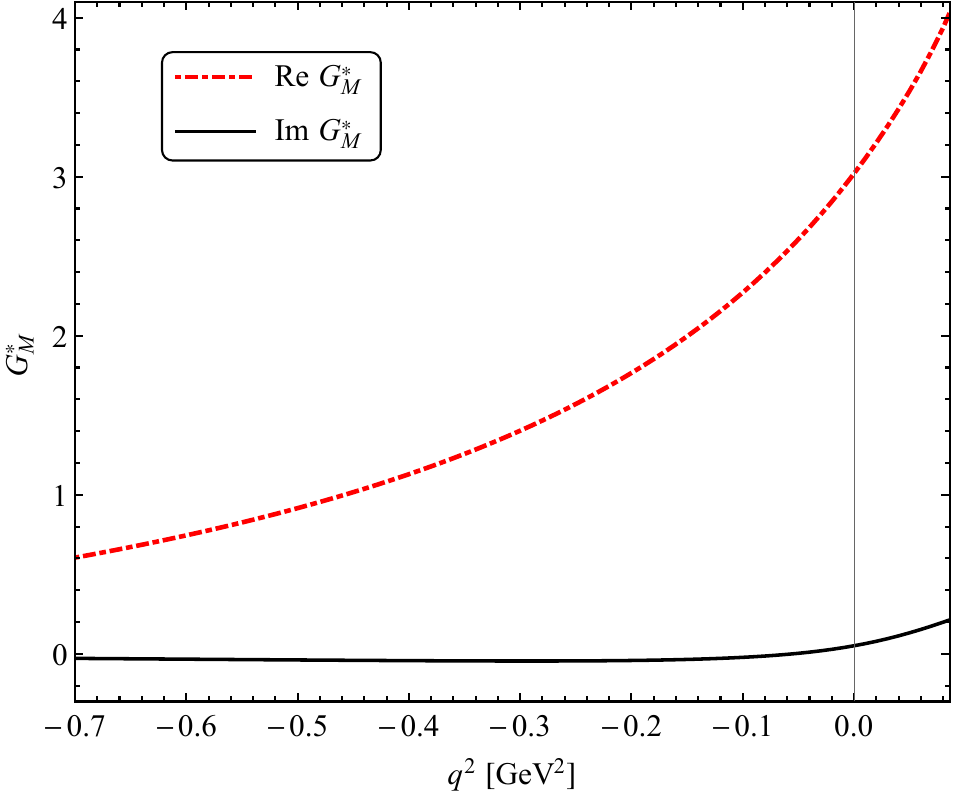}
    \caption{Real (red) and imaginary (black) part of the magnetic TFF in the space- and time-like region, $- 0.7\, \text{GeV}^2 < q^2 < (m_\Delta - m_n)^2$. The curves correspond to the red curve in Fig.\ \ref{fig:ImM-various-signs} obtained using $h_A > 0$ and the corresponding fit result $ P_{M}^* = 406.12\,$GeV$^{-2}$ with the cutoff at $\Lambda = 2\,$GeV. The value Im$G^*_M(0) =  0.0503$ is obtained from an unsubtracted dispersion relation.}
    	\label{fig:GMspacelike}
\end{figure}

In Fig.\ \ref{fig:GMspacelike} we show real and imaginary part of $G_M^*$ for space- and time-like $q^2$. We have limited the latter region to the physical decay region where the Dalitz decay can take place. We stress that $G_M^*(q^2)$ is {\em not} purely real in the spacelike region $q^2 < 0$. While the shape of the imaginary part is a prediction of our theory, one has to take the overall size, i.e.\ the value at the photon point, with a grain of salt. A subtracted dispersion relation cannot predict Im$G_M^*(0)$, while the measurements of $\gamma N \to \Delta$ provide only the real part at the photon point. We use here the result from an unsubtracted dispersion relation. To judge the quality of this approximation, we provide also the value of Re$G_M^*(0)$ as obtained from the unsubtracted dispersion relation, but using the value $P_M^*=406.12\,$GeV$^{-2}$ coming from a fit of the subtracted dispersion relation to the data; see Fig.\ \ref{fig:ImM-various-signs}. In this way we find $G_M^*(0) = 3.3126 + 0.0503 i$. The real part is only 10\% away from the experimental value. This provides some support that the value for the imaginary part is reasonable. In particular, it suggests that the imaginary part is small (but not zero) in the spacelike region.  

In the timelike region, Fig.\ \ref{fig:GMspacelike} shows that the real part grows from about 3 at the photon point to about 4 at the end of the Dalitz decay region. This variation is comparable but somewhat larger than the result of the quark-model calculation of \cite{Ramalho:2015qna}. But it also shows that one needs a resolution of 30\% or better to distinguish in the Dalitz decay region the true form factor from a constant. 

In Table \ref{table:radii-res}, we display FF value (magnetic moment) and slope (radius) at the photon point and explore the impact of parameter variations. Note that since Re$G_M^*(0)$ is fitted to the data, it cannot change. The imaginary part, however, obtained from an unsubtracted dispersion relation, could change but it remains the same within the accuracy that we display. 
\begin{table}
	\centering	
	\begin{tabular}{|c||c|c|}
          \hline 
          & $ G^{*}_{M} (0)$ & $ \langle r^2 \rangle^*_{M}$ [GeV$^{-2}$]  \\
          \hline 
          \hline 
          $\Lambda = 1\,$GeV & $3.02 + 0.05\,i$ & $19.03 + 2.23\,i$ \\
          \hline 
          $\Lambda = 2\,$GeV & $3.02 + 0.05\,i$ & $18.35 + 2.19\,i$ \\
          \hline
          $h_A \approx 2.9$ & $3.02 + 0.05\,i$ & $18.76 + 2.43\,i$ \\
          \hline
          $h_A \approx 2.4$ & $3.02 + 0.05\,i$ & $17.95 + 1.94\,i$ \\
          \hline
          $ H_A \approx 2.5$ & $3.02 + 0.05\,i$ & $18.66 + 2.19\,i$ \\
          \hline
          $H_A \approx 2.0$ & $3.02 + 0.05\,i$ & $18.05 + 2.19\,i$ \\
          \hline 
      \end{tabular}
	\caption{Sensitivity of magnetic transition moment and magnetic radius with respect to parameter variations. We use $\Lambda = 2\,$GeV and central values for $h_A$ and $H_A$ unless otherwise stated.}
	\label{table:radii-res}
\end{table}

More interesting are the results for the radius. They are a pure prediction of our subtracted dispersion relations. First of all, we observe that the non-vanishing imaginary part has grown to a 10\% effect, while it is less than 2\% for the magnetic moment. It remains to be seen if there is a way how to verify this effect experimentally. Presumably one has to go away from the peak position of the Delta and carry out a full-fledged analysis of the variation in the invariant mass of the pion-nucleon system that emerges from the produced Delta; see also the discussion in Appendix \ref{sec:exp-re-BW}. We have already stressed in the introductory section that this is far beyond the scope of the present paper. But the non-trivial and quantitatively non-negligible imaginary part of the radius is certainly an intriguing result.

The sensitivity to parameter variations is mild, the largest impact is caused by a change in the cutoff. The size of the radius is completely reasonable. A size of $0.85\,$fm would correspond to a squared radius of about $19\,$GeV$^{-2}$. Increasing one of the three-point coupling constants, i.e.\ increasing the importance of the nucleon or Delta exchange diagram, leads to an increase of the radius. Thus all effects --- nucleon exchange, Delta exchange (our proxy for correlated pion-nucleon pair exchange) and contact term (the short-distance physics) --- add up constructively for the magnetic radius. 
Of course, the radius can be addressed from the spacelike and the timelike side. For the latter we have now full predictive power.

\subsection{Dalitz decay distribution}
\label{sec:res-Dalitz}

Given that the smaller form factors contribute only on the sub-percent level, we can predict the Dalitz decay $\Delta \to n \; e^+ e^-$ based on the magnetic dipole TFF. The same idea has been utilized in \cite{Ramalho:2015qna} and has entered the corresponding discussion in the experimental paper \cite{Adamczewski-Musch:2017hmp} by the HADES collaboration.

\begin{figure}
  \centering
    \includegraphics[width=0.6\textwidth]{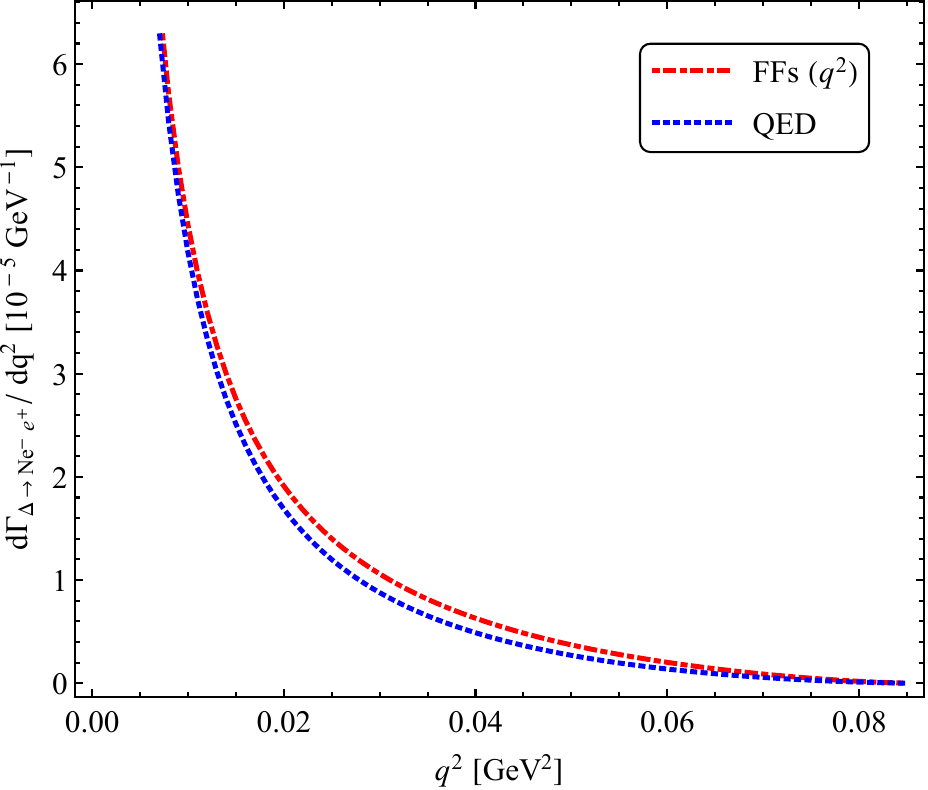}
  \caption{Single-differential decay width for the $\Delta\rightarrow N \, e^+ e^-$ Dalitz decay. The top curve, labeled ``FFs$\,(q^2)$'', is the angular integral of \eqref{eq:Dalitzdistr}. The bottom curve is the QED analogue, given by \eqref{eq:QEDcase}.}
  \label{fig:Gamma}
\end{figure}

In Fig.\ \ref{fig:Gamma} we plot both the single differential decay width $\mathrm{d}\Gamma/\mathrm{d}q^2$ for the Dalitz decay $\Delta\rightarrow n e^+ e^-$ and the corresponding QED case \eqref{eq:QEDcase}, in the kinematically allowed range $4m_e^2 < q^2 < (m_{\Delta}-m_n)^2$. The first is the angular integral of \eqref{eq:Dalitzdistr} or its approximation \eqref{eq:Dalitzdistr2}, the second is a simplified version, blind to the $q^2$-dependence of the TFF. 
The discrepancy between these two curves is caused by the non-trivial internal structure of baryons and is therefore of great interest.

What could already be anticipated from Fig.\ \ref{fig:GMspacelike}, given the limited variation in the timelike region, can now also be seen fully quantitatively in Fig.\ \ref{fig:Gamma}. In order to resolve the difference between a non-trivial TFF and the pointlike (``QED'') case one needs a rather high resolution in the tail of the Dalitz distribution. 
At present, the experimental accuracy of the HADES experiment \cite{Adamczewski-Musch:2017hmp} is not high enough to resolve this
difference. But our plot shows the accuracy that is required. 
Variations of our parameters according to \eqref{eq:param-var-hA-HA} lead to changes that lie within the line width of the ``FFs'' curve of Fig.\ \ref{fig:Gamma}.

\section{Extended outlook}
\label{sec:outlook}

Given the success of our model-independent framework based on dispersion theory and $\chi$PT concerning the numerically dominant piece, the magnetic dipole TFF, it is interesting to check how well our formalism is doing for the two smaller (quadrupole) TFFs. Yet we hesitate to provide a full-fledged analysis for the reasons spelled out previously. In particular, our precision is limited by the following approximations: treating the Delta as if it was a stable asymptotic state; using the Delta exchange diagram as a proxy for pion-nucleon rescattering; using dispersion relations for Jones-Scadron FFs instead of constraint-free FFs. Nonetheless, to satisfy our own curiosity we have carried out an analysis of one of the smaller TFFs, namely the electric quadrupole. It shares with the magnetic dipole the property that the subtraction constant $G(0)$ can be determined at the photon point from the decay $\Delta \to N \gamma$. The same is not true for the Coulomb quadrupole, which does not enter the decay formula \eqref{eq:photCase}. The reason is simply the fact that a real photon cannot have vanishing helicity. Therefore $G_C^* \sim G_0$ cannot be populated for real photons.

\begin{figure}
    \centering
     \includegraphics[width=0.6\textwidth]{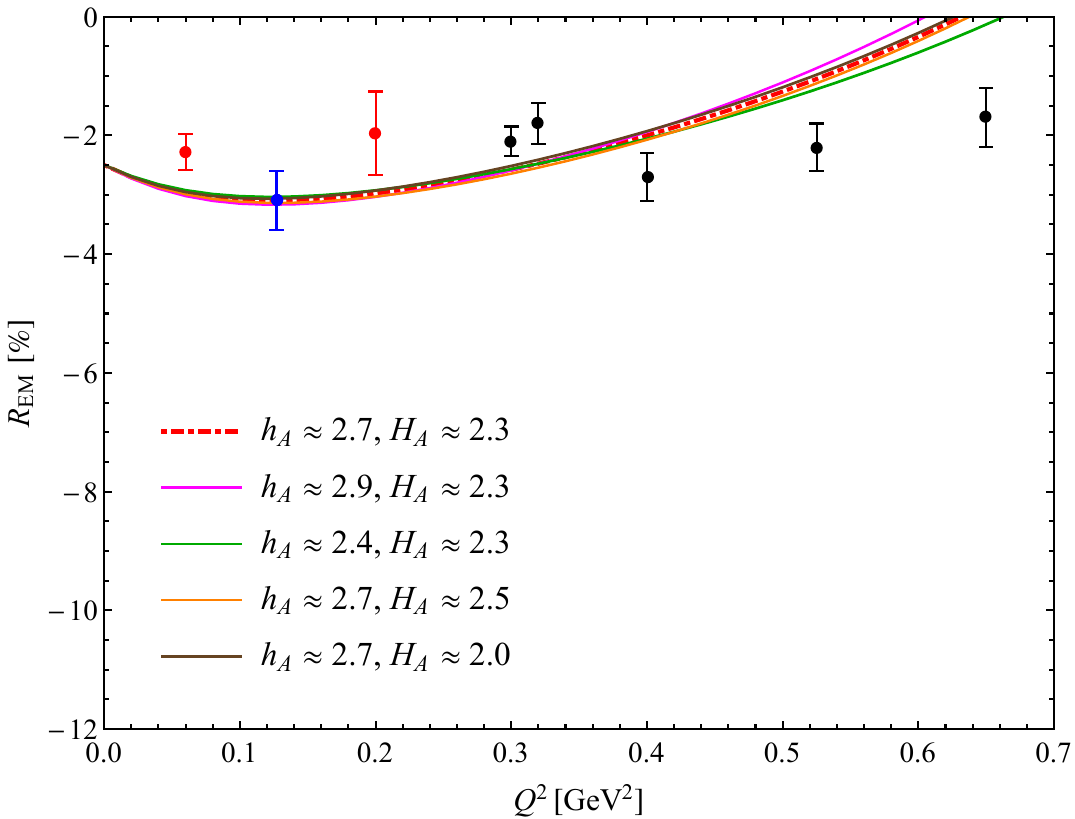}
    \caption{The ratio $R_{\rm EM}$ as a function of $Q^2=-q^2$. The various lines display the impact of parameter variations. To the right from top to bottom:
    $h_A$ increased by 10\%, central value for $H_A$ (magenta); 
    $H_A$ decreased by 10\%, central value for $h_A$ (brown);
    central values for $h_A$ and $H_A$ (red, dash-dotted); 
    $H_A$ increased by 10\%, central value for $h_A$ (orange);
    $h_A$ decreased by 10\%, central value for $H_A$ (green). For the color code of the data see the figure caption of Fig.\ \ref{fig:ImM-various-signs}.} 
    	\label{fig:REM}
\end{figure}

We use the subtracted dispersion relation for $G_E^*(-Q^2)$ and our previous result for $G^*_M(-Q^2)$ to obtain $R_{\rm EM}$ from \eqref{eq:multip-ratios}, using $G_E^*(0)$ and $P_E^*$ as fit parameters. The results are provided in Fig.\ \ref{fig:REM}. 
For the central values of $h_A$ and $H_A$ we find $P_E^* \approx 10.24\,\text{GeV}^{-2}$. We observe a fair agreement with the data up to about $Q^2 \approx 0.5\,$GeV$^{2}$, even though the quality is not as impressive as for the magnetic case displayed in Fig.\ \ref{fig:hAvariation}. Variation of the input parameters is also displayed in Fig.\ \ref{fig:REM} by the various lines. All lines lie close to each other, indicating again a very robust result. The next thing to notice is the qualitative feature of a dip at around $Q^2 \approx 0.12\,$GeV$^2$. Interestingly, the data do not really show this dip, but do not exclude it either. On the other hand, theoretical considerations indicate that this dip is caused by pion-cloud effects \cite{Pascalutsa:2007wz,Friedrich:2003iz}. Of course, the pion-cloud physics is covered by our dispersive framework, which features the fact that at low energies virtual photons couple dominantly to the lightest degrees of freedom, the pions as the Goldstone bosons of chiral symmetry breaking. 

\begin{figure}
    \centering
     \includegraphics[width=0.6\textwidth]{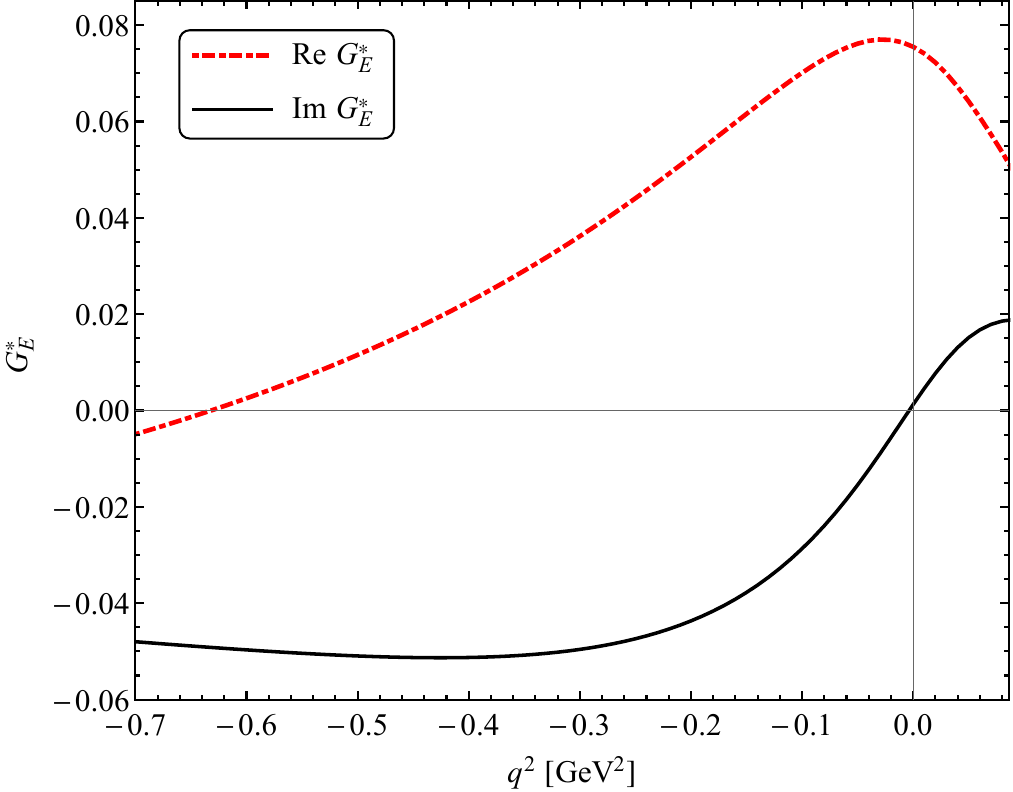}
    \caption{Same as Fig.\ \ref{fig:GMspacelike} but for the electric TFF. The value of the contact term is $P_{E}^* = 10.24\,$GeV$^{-2}$.}
    	\label{fig:GEspacetimelike}
\end{figure}

In Fig.\ \ref{fig:GEspacetimelike} we provide the corresponding plot for $G_E^*(q^2)$ itself, using central parameter values. A remarkable feature is a maximum of the real part, which lies very close to the photon point. As a consequence, close to the photon point the shape of the ratio $R_{\rm EM}$ is dominated by the change of $G_M^*$, leading to a drop of $R_{\rm EM}$ in Fig.\ \ref{fig:REM}. When moving further into the spacelike region, $G_E^*(q^2)$ develops a slope, eventually a larger one than $G_M^*$. This leads to a rise in the ratio $R_{\rm EM}$.  

An extremum close to the photon point seems to be a generic feature of the smaller FFs of the Delta-to-nucleon transition. In the advanced data parametrization of Ref.\ \cite{Eichmann:2018ytt}, taking into account all kinematical constraints, one sees the occurrence of extrema in all helicity amplitudes. Only the particular combination of helicity amplitudes \eqref{eq:Gs-jones-scadron} that lead to the magnetic dipole TFF, Fig.\ \ref{fig:GMspacelike}, does not show this extremum. 

It is also interesting to observe that our results for the real and imaginary part of $G_E^*(q^2)$ have about the same size in the spacelike region. This is in sharp contrast to the magnetic case. The overall picture is that in the spacelike region the real part of the magnetic dipole TFF is large, while all other quantities (the imaginary part of the magnetic TFF and both real and imaginary part of the electric TFF) are small in magnitude, but of comparable absolute size. 

This has interesting consequences for the radius, provided in Table \ref{table:radii-res-E}. 
\begin{table}
	\centering	
	\begin{tabular}{|c||c|c|}
          \hline 
          Quantity & $ G^{*}_{E} (0)$ & $ \langle r^2 \rangle^*_{E}$ [GeV$^{-2}$]   \\
          \hline \hline	
          $\Lambda = 1\,$GeV & $0.07 + 0.00\,i$ & $- 8.01 + 28.08\,i$ \\
          \hline
          $\Lambda = 2\,$GeV & $0.07 + 0.00\,i$ & $- 8.69 + 27.02\,i$ \\
          \hline
          $h_A \approx 2.9$ & $0.07 + 0.00 \,i$ & $- 11.01 + 29.75\,i$ \\
          \hline
          $h_A \approx 2.4$ & $0.07 + 0.00 \,i$ & $- 6.38 + 24.30\,i$ \\
          \hline
          $ H_A \approx 2.5$ & $0.07 + 0.00 \,i$ & $- 9.18 + 27.03 \,i$ \\
          \hline
          $H_A \approx 2.0$ & $0.07 + 0.00\,i$ & $- 8.21 + 27.02\,i$ \\
          \hline	
        \end{tabular}
	\caption{Same as Table \ref{table:radii-res} but for the electric quantities.}
	\label{table:radii-res-E}
\end{table}
While for the magnetic case, Table \ref{table:radii-res}, the respective real parts are always dominant relative to the imaginary parts, our formalism produces a very large imaginary part for the electric quadrupole radius. While parameter variations have quite some impact, the overall qualitative result is robust. At present, it is not clear if this is a physical effect or a deficiency of our approach. After all, we address here very small quantities, relative to the magnetic TFF. This requires high theoretical accuracy. If the large imaginary part is a true physical effect, it will be interesting to figure out how one can explore it experimentally.

Overall, we regard our results for the magnetic TFF as solid predictions, but interpret our results for the much smaller electric TFF as a motivation for further studies and further improvements of the formalism. 

\vspace*{2em}

{\bf Acknowledgements:} This work has been supported by the Swedish Research Council (Vetenskapsr\aa det) (grant number 2019-04303)
and by Thailand International Development Cooperation Agency (TICA), Thailand Science Research and Innovation (TSRI), National Research Council of Thailand (NRCT) from the Royal Golden Jubilee (RGJ) Program, Center of Excellence in High Energy Physics and Astrophysics (CoE) at Suranaree University of Technology, Swedish International Development Cooperation Agency (Sida) and International Science Programme (ISP) at Uppsala University under the Thai-Swedish Trilateral Development Cooperation Programme (contract no.\ PHD/0103/2560).

\appendix

\section{Which quantities are really measured in the spacelike region?}
\label{sec:exp-re-BW}

Experimentally the TFFs are extracted from the reaction $e^- N \to e^- \Delta \to e^- \, \pi N$ in the 
one-photon approximation, i.e.\ one studies formally the reaction $\gamma^* N \to \Delta \to \pi N$. Which events from all 
final-state $\pi N$ pairs are used to reconstruct the $\Delta$? Clearly, if one plots the number of events as a function of the 
invariant mass $m_{\pi N}$ of the final-state $\pi N$ pair, one observes a relatively broad bump with width $\Gamma_\Delta$ 
that peaks at $m_\Delta$. 
In principle, there are two possibilities how to continue: \\ a) One could take all events in the bump region, 
e.g.\ $m_\Delta - \Gamma_\Delta/2 < m_{\pi N} < m_\Delta - \Gamma_\Delta/2$ (and subtract the background below the bump);\footnote{This is what is done in the timelike region of Dalitz decays \cite{Adamczewski-Musch:2017hmp}.}  \\ 
or b) one could take just the events from the bin 
at the peak position, 
i.e.\ $m_{\pi N} \approx m_\Delta$. \\ 
It turns out that the second option is used by the experimentalists. This is spelled out explicitly, e.g., 
in \cite{Pascalutsa:2006up} between equations (2.9) and (2.10). 

In the following, we determine how the scattering amplitudes for $\gamma^* N \to \Delta \to \pi N$ with $m_{\pi N} \approx m_\Delta$
are related to our TFFs. In that way we are aiming at the justification of \eqref{eq:GM-iMM}. Why is the real part used on the left-hand side? And not, e.g., the modulus? 
We stick to a one-loop calculation and treat vertices, in particular 
their spinor structure, very schematically. The quantities ${\cal F}$, ${\cal T}$ and ${\cal P}$ 
that we introduce in the following are supposed to have no 
imaginary parts in the kinematical region that we study. 

The ``initial'' $\gamma^* N$ system can couple to the $\Delta$ directly or via a $\pi N$ loop. We denote the direct coupling by 
${\cal T}$ and the four-point $\gamma^* N$-$\pi N$ structure by ${\cal F}$. The three-point coupling $\Delta$-$\pi N$ is denoted 
by ${\cal P}$. Finally, the $\pi N$ loop is given by
\begin{eqnarray}
  \label{eq:def-loop}
  L := \int \frac{d^4q}{(2\pi)^4} \, \frac{1}{q^2-m_\pi^2 + i \epsilon} \, \frac{1}{(p_\Delta-q)^2 - m_N^2 + i \epsilon}  \,.
\end{eqnarray}
Here $p_\Delta$ is the $\Delta$ momentum, i.e.\ the total momentum of the final $\pi N$ pair. 
Using retarded propagators and/or Wick rotation, one can show 
\begin{eqnarray}
  \label{eq:loop2}
  {\rm Im}(iL) = -\frac{p_{\rm cm}}{8 \pi m_{\pi N}}  
\end{eqnarray}
with the momentum of $\pi$ and $N$ in their center-of-mass frame, 
\begin{eqnarray}
  \label{eq:defpcmDelta}
  p_{\rm cm} = \frac{1}{2 m_{\pi N}} \, \lambda^{1/2}(m_{\pi N}^2,m_\pi^2,m_N^2)  \,.
\end{eqnarray}
Note that at the peak position where $m_{\pi N} = m_\Delta$ one has $p_{\rm cm} = k_\Delta$ with the latter being introduced after 
\eqref{eq:GM-iMM}. 

A TFF, $G$, is obtained by 
\begin{eqnarray}
  \label{eq:calcG-schematic}
  i G = i {\cal T} + i {\cal F} \, i^2 L \, i {\cal P} 
\end{eqnarray}
which leads to
\begin{eqnarray}
  \label{eq:calcG-schematic2}
  G = {\cal T} - {\cal F} \, iL \, {\cal P} \,. 
\end{eqnarray}
For later use we determine
\begin{eqnarray}
  \label{eq:calcG-schematicRe}
  {\rm Re}G = {\cal T} - {\cal F} \, {\rm Re}(iL) \, {\cal P} \,. 
\end{eqnarray}

The Feynman matrix element ${\cal M}$ for the reaction $\gamma^* N \to \Delta \to \pi N$ is obtained as
\begin{eqnarray}
  \label{eq:calcM-schematic}
  i {\cal M} = i {\cal T} \, i S \, i {\cal P} + i {\cal F} \, i^2 L \, i {\cal P} \, i S \, i {\cal P} \,.  
\end{eqnarray}
Here $S$ denotes the full $\Delta$ propagator
\begin{eqnarray}
  \label{eq:defSDelta}
  S = \frac{1}{p_\Delta^2 - m_\Delta^2 - \Pi} \approx \frac{1}{p_\Delta^2 - m_\Delta^2 - i {\rm Im}\Pi} 
\end{eqnarray}
with the self-energy $\Pi$ \cite{pesschr} given by 
\begin{eqnarray}
  \label{eq:calcselfen}
  -i \Pi = i {\cal P} \, i^2 L \, i{\cal P} = {\cal P}^2 L  \,.
\end{eqnarray}
For later use (we are aiming at a justification of \eqref{eq:GM-iMM}) we calculate
\begin{eqnarray}
  \label{eq:calcM-schematic-Im}
  {\rm Im}{\cal M} &=& - {\cal T} \, {\cal P} \, {\rm Im}S + {\cal F} \, {\cal P}^2 \, {\rm Im}(i L \, S)  \nonumber \\
  &=& - {\cal T} \, {\cal P} \, {\rm Im}S + {\cal F} \, {\cal P}^2 \, {\rm Re}(iL) \, {\rm Im}S  \nonumber \\
  && {} + {\cal F} \, {\cal P}^2 \, {\rm Im}(iL) \, {\rm Re}S \,.
\end{eqnarray}
Up to normalization issues, ${\cal M}$ should agree with $M_{1+}^{(3/2)}$, which appears in \eqref{eq:GM-iMM}. 

The width of the $\Delta$ is practically given by its decay to $\pi N$. With the Breit-Wigner formula
\begin{eqnarray}
  \label{eq:SDelta-BW}
  S \approx \frac{1}{p_\Delta^2 - m_\Delta^2 + i m_\Delta \Gamma_\Delta} 
\end{eqnarray}
this implies 
\begin{eqnarray}
  \label{eq:width-selfen}
  m_\Delta \Gamma_\Delta = - {\rm Im}\Pi = - {\cal P}^2 \, {\rm Im}(iL)  \,.
\end{eqnarray}

Recall that $p_\Delta$ denotes the total momentum of the final $\pi N$ system; thus $p_\Delta^2 = m_{\pi N}^2$. 
So far we have not specified $m_{\pi N}$. But following the experimental procedure, the choice $m_{\pi N} = m_\Delta$ leads to 
\begin{eqnarray}
  \label{eq:reimS}
  {\rm Re}S = 0 \,, \qquad {\rm Im}S = -\frac1{m_\Delta \Gamma_\Delta}  \,.
\end{eqnarray}
For this choice of the two-body mass, i.e.\ at the peak position of the $\Delta$ bump, 
equation \eqref{eq:calcM-schematic-Im} simplifies to 
\begin{eqnarray}
  \label{eq:calcM-schematic-Im2}
  {\rm Im}{\cal M} &=& - {\cal T} \, {\cal P} \, {\rm Im}S + {\cal F} \, {\cal P}^2 \, {\rm Re}(iL) \, {\rm Im}S \nonumber \\
  &=& \left(-{\cal T} + {\cal F} \, {\rm Re}(iL) \, {\cal P} \right) \, {\cal P} \, {\rm Im}S  \nonumber \\
  &=& - {\rm Re}G \, {\cal P} \, {\rm Im}S  
\end{eqnarray}
where we have used \eqref{eq:calcG-schematicRe} in the last step. 
We can invert this equation and use \eqref{eq:reimS}, \eqref{eq:width-selfen}, and \eqref{eq:loop2} to obtain
\begin{eqnarray}
  \label{eq:justify13}
  {\rm Re}G &=& -\frac{{\rm Im}{\cal M}}{{\cal P} \, {\rm Im}S} = \frac{{\rm Im}{\cal M} \, m_\Delta \Gamma_\Delta}{{\cal P}}  \nonumber \\
  &\sim & {\rm Im}{\cal M} \, m_\Delta \Gamma_\Delta \, \sqrt{\frac{-{\rm Im}(iL)}{m_\Delta \Gamma_\Delta}}
  \sim {\rm Im}{\cal M} \, \sqrt{k_\Delta \, \Gamma_\Delta}  \,.
\end{eqnarray}
This last version compares well with \eqref{eq:GM-iMM} concerning the appearance of the $\Delta$ width and the phase-space factor
$\sim k_\Delta$. Equation \eqref{eq:justify13} specifies that it should be the real part of the TFF that 
appears on the left-hand side of \eqref{eq:GM-iMM}. 
The procedure to use the real parts is in line with \cite{Hilt:2017iup}.

\section{Contributions from the anomalous cut}
\label{sec:anom-cutt-appendix}

\subsection{General considerations about the analytic structure}
\label{subsec:genana}

The extra term $T^{\rm anom}(s)$ in (\ref{eq:tmandel}) arises if 
the mass $m_{\rm exch}$ of the exchanged state in a crossed channel is sufficiently  light to fulfill the anomalous threshold condition \cite{Karplus:1958zz}:
\begin{eqnarray}
  \label{eq:anomthr}
  m_{\rm exch}^2 < \frac12 \, \left(m_{\Delta}^2 + m_n^2 - 2 m_\pi^2 \right)  \,.
\end{eqnarray}
For the formal reaction $\Delta^0 \bar{n} \to \pi^+ \pi^-$, one has to consider the exchange of the proton and $\Delta^\pm$ resonances. 
The condition \eqref{eq:anomthr} does not hold for the $\Delta$ exchange, but is satisfied for the proton exchange. In this case, the 
logarithm obtained from the partial-wave projection (\ref{eq:def-redampl})  
has a cut in the complex $s$ plane that intersects with the unitarity cut. Part of this cut lies on the physical Riemann sheet. Dispersion relations have to be modified accordingly in order to produce the correct results.  In practice, the contour has to circumnavigate the new cut in addition to the familiar two-pion cut. As explored in \cite{Junker:2019vvy}, a dispersive representation that ignores such anomalous cut produces incomplete results.
To disentangle the cuts, 
one can define the cut of the logarithm such that it connects the branch point to the unitarity cut by a straight
line. The additional contribution $T^{\rm anom}(s)$ takes care of the extra cut. 

To be more concrete, we note that the p-wave partial-wave projection of type (\ref{eq:def-redampl}) for a $t$- or $u$-channel
exchange process produces a term 
\begin{eqnarray}
  K(s) &=& g(s) - \frac{2 f(s)}{Y(s) \, \kappa^2(s)} \nonumber \\
  &&{}+ f(s) \, \frac{1}{\kappa^3(s)} \, \log\frac{Y(s)+\kappa(s)}{Y(s)-\kappa(s)}
  \label{eq:Klog}
\end{eqnarray}
with the functions $Y$, $\kappa$ and $\sigma$ defined as
\begin{eqnarray}
  \label{eq:defY}
  Y(s) := s + 2 m_{\rm exch}^2 - m_1^2 - m_2^2 - 2 m_\pi^2  \,,
\end{eqnarray}
\begin{eqnarray}
  \label{eq:defKacser}
  \kappa(s) := \lambda^{1/2}(s,m_1^2,m_2^2) \, \sigma(s) \,,
\end{eqnarray}
and 
\begin{eqnarray}
  \label{eq:defsigma}
  \sigma(s) := \sqrt{1-\frac{4m_\pi^2 }{s}}  \,.
\end{eqnarray}
 Here we choose
$m_1 = m_{\Delta}$, $m_2 = m_n$.
Specific formulae for the functions $f(s)$, $g(s)$  can be deduced from the matrix elements provided in Section \ref{sec:chiPT}.
Obviously Eq.\ \eqref{eq:Klog} is ill-defined for $Y(s)=0$. When the exchanged baryon is the $\Delta$ resonance, this point lies outside the integration path, bringing no additional complication to our formalism. 
However, in the case of proton exchange, this point is located on the unitarity cut, at $s_Y=0.675$ GeV$^2$. Therefore, the logarithm needs an analytic continuation along the unitarity cut. For convenience one can consider different intervals delimited by the 
following four points: 
\begin{itemize}
\item At the scattering threshold $s_{st}:=(m_{\Delta}+m_n)^2$ we have $\kappa=0$. 
  Above this point, i.e.\ for $s$ real and larger than $s_{st}$, there is the true scattering 
  region. There, $\kappa$ is real and $Y$ is 
  positive and larger than $\kappa$. The logarithm in (\ref{eq:Klog}) can be defined as the real-valued standard logarithm of 
  positive numbers. 
\item At $s_Y:=m_{\Delta}^2 + m_n^2 + 2 m_\pi^2-2 m_{p}^2$ we have $Y=0$. 
  For $s$ real and between $s_Y$ and $s_{st}$ the function $\kappa$ is purely imaginary and
  $Y$ is still positive.
\item At the decay threshold\footnote{Concerning the phrase ``decay threshold'' we note that for $s<(m_\Delta-m_n)^2$ the decay $\Delta \to n \, 2\pi$ is possible.} 
$s_{dt}:=(m_{\Delta}-m_n)^2$ we have $\kappa=0$. For $s$ real and between $s_{dt}$ and $s_Y$ the function $\kappa$ is 
  purely imaginary and $Y$ is negative. 
\item At $s_{2\pi}:= 4 m_\pi^2$  we have $\kappa=0$. For $s$ real and between $s_{2\pi}$ and $s_{dt}$ the function $\kappa$ is real
  and $Y$ is negative.
\end{itemize}
When the exchanged baryon is the proton we have $0< s_{2\pi} < s_{dt} < s_Y < s_{st}$. 
The function $K$ in (\ref{eq:Klog}) that enters finally (\ref{eq:tmandel}) is then defined on the relevant part of the 
real axis by
\begin{equation}
  \label{eq:properdefKl4}
  K(s) := g(s) - \frac{2 f(s)}{Y(s) \, \kappa^2(s)} + \frac{f(s)}{\kappa^3(s)} \, \log\frac{Y(s)+\kappa(s)}{Y(s)-\kappa(s)}
\end{equation}
for $s > s_{st}$, by 
\begin{equation}
  \label{eq:properdefK34}
  K(s) := g(s)
  - \frac{2 f(s)}{Y(s) \, \kappa^2(s)} + \frac{2f(s)}{\kappa^2(s) \, \vert\kappa(s)\vert} \arctan\frac{\vert\kappa(s)\vert}{Y(s)} 
\end{equation}
for $s_Y < s < s_{st}$, by 
\begin{eqnarray}
  K(s) &:=& g(s) - \frac{2 f(s)}{Y(s) \, \kappa^2(s)}  \nonumber \\
  && {}+ \frac{2f(s)}{\kappa^2(s) \, \vert\kappa(s)\vert} \left( \arctan\frac{\vert\kappa(s)\vert}{Y(s)} +\pi \right) 
  \label{eq:properdefK23}
\end{eqnarray}
for $s_{dt} < s < s_Y$, and by 
\begin{equation}
  \label{eq:properdefK12}
  K(s) := g(s) - \frac{2 f(s)}{Y(s) \, \kappa^2(s)} + \frac{f(s)}{\kappa^3(s)} \, \left(\log\frac{Y(s)+\kappa(s)}{Y(s)-\kappa(s)}+2 i \pi\right)
\end{equation}
for $s_{2\pi}< s <s_{dt}$.
Here the standard logarithm for positive real numbers is used and the arctan function with values between $-\pi/2$ and $\pi/2$.

The extra $\pi$ in (\ref{eq:properdefK23}) and $2 i \pi$ in (\ref{eq:properdefK12}) ensure a smooth analytic continuation of $K$. However, such an extra piece creates a singularity where $\kappa$ vanishes. This happens at the decay threshold $s_{dt}$; see also the discussion in \cite{Schneider:2012ez} and references therein. In principle, the singularity does not carry over to the FF and is of no physical significance. But it is numerically unpleasant to deal with explicitly. Therefore we have decided to modify the integration contour and avoid the singularity altogether. This is better explained after finishing the discussion of the analytic structure of our amplitudes. We will come back to the singularity in Subsection \ref{sec:anom-num}. 

The branch points of the logarithm in (\ref{eq:Klog}) satisfy 
$Y^2(s)=\kappa^2(s)$. They are located at 
\begin{eqnarray}
  s_\pm &=& -\frac12 \, m_{\rm exch}^2 + \frac12 \left(m_{\Delta}^2 + m_n^2 + 2 m_\pi^2 \right) \nonumber \\ && {}
  - \frac{m_{\Delta}^2 \, m_n^2 - m_\pi^2 \, (m_{\Delta}^2 + m_n^2) + m_\pi^4}{2m_{\rm exch}^2} 
  \nonumber \\ && {}
  \mp \frac{\lambda^{1/2}(m_{\Delta}^2,m_{\rm exch}^2,m_\pi^2) \, \lambda^{1/2}(m_{\rm exch}^2,m_n^2,m_\pi^2)}{2m_{\rm exch}^2} \,.
  \phantom{mm}
  \label{eq:defspm}
\end{eqnarray}
We take $s_+$ as the solution that has a positive imaginary part for small values of $m_{\Delta}^2$.  For the case $m_{\rm exch}^2= m_p^2$, the trajectory of $s_+$ as a function of  $m_{\Delta}^2 +i\epsilon$ 
intersects with the unitarity cut where (\ref{eq:anomthr}) 
turns to an equality. In particular, for the physical value of $m_{\Delta}^2$, the branch point  $s_+$ is located in the lower half plane of the first Riemann sheet:
\begin{eqnarray}
  s_+ &=& -\frac12 \, m_{p}^2 + \frac12 \left(m_{\Delta}^2 + m_n^2 + 2 m_\pi^2 \right) \nonumber \\ && {}
  - \frac{m_{\Delta}^2 \, m_n^2 - m_\pi^2 \, (m_{\Delta}^2 + m_n^2) + m_\pi^4}{2m_{p}^2} 
  \nonumber \\ && {}
  -i \, \frac{\lambda^{1/2}(m_{\Delta}^2,m_{p}^2,m_\pi^2) \, \left(-\lambda(m_{p}^2,m_n^2,m_\pi^2)\right)^{1/2}}{2m_{p}^2} 
  \phantom{mm}
  \label{eq:defspconcr}
\end{eqnarray}
with positive square roots.
This is the starting point for the definition of the anomalous contribution $T^{\rm anom}$ that enters \eqref{eq:tmandel}:
\begin{eqnarray}
  T^{\rm anom}(s) &=& \Omega(s) \, s \, \int\limits_0^1 \text{d}x \, \frac{\text{d}s'(x)}{\text{d}x}  \,
                      \frac{1}{s'(x)-s}  \nonumber \\
                  && \times 
                     \frac{2f(s'(x))}{(-\lambda(s'(x),m_{\Delta}^2,m_n^2))^{1/2} \, \kappa^2(s'(x))}  \nonumber \\
                  && \times  \frac{t(s'(x))}{\Omega(s'(x)) \, s'(x)} \phantom{mm}
  \label{eq:tanom}
\end{eqnarray}
with the straight-line path
\begin{eqnarray}
  \label{eq:defsx}
  s'(x) := (1-x) s_+ + x\, s_c
\end{eqnarray}
that connects the branch point (\ref{eq:defspconcr}) of the logarithm in (\ref{eq:Klog}) to the point $s_c=5m_\pi^2$ on the unitarity cut. We have dedicated Subsection \ref{sec:anom-num} below to motivate why we chose to connect the branch point $s_+$ to the point $s_c$, instead of to the two-pion threshold point $s_{2\pi}$, as originally done in \cite{Junker:2019vvy}. But before turning to this issue, we specify the further ingredients.

An analytic continuation of the scattering amplitude $t(s)$ in the complex plane is needed for the anomalous part of Eq.\ \eqref{eq:tanom}.
We take from \cite{Dax:2018rvs} the following expressions (extended to the complex plane).
The approximation from $\chi$PT is given by
\begin{equation}
\label{tChPT1}
t_{\chi\rm PT}(s) \approx t_2(s)+t_4(s)
\end{equation}
and its unitarized version is
\begin{equation}
\label{tIAM1}
t_{\text{IAM}}(s)=\frac{t_2^2(s)}{t_2(s)-t_4(s)} 
\end{equation}
with
\begin{eqnarray}
  t_2(s) &=& \frac{s\sigma^2}{96\pi F_0^2}  \,, 
\end{eqnarray}
\begin{widetext}
\begin{eqnarray}
  t_4(s) &=& \frac{t_{2}(s)}{48\pi^2F_0^2}\bigg[s\left(\bar{l}+\frac{1}{3}\right)-\frac{15}{2} m_\pi^2
             -\frac{m_\pi^4}{2s}\Big(41-2L_\sigma\big(73-25\sigma^2\big)
             +3L_\sigma^2\big(5-32\sigma^2+3\sigma^4\big) \Big)\bigg] 
             \nonumber \\[0.7em] && {}-\hat\sigma(s) \,t_2^2(s)   \,,
            \label{t2t4}
\end{eqnarray}
\end{widetext}
\begin{equation}
  \label{Abbrev_Lsig_sig}
  L_\sigma=\frac{1}{\sigma^2}\left(\frac{1}{2\sigma}\log\frac{1+\sigma}{1-\sigma}-1\right)   \,.
\end{equation}
The function $\sigma(s)$ is defined in \eqref{eq:defsigma}   and $\hat\sigma(s)$ by
\begin{eqnarray}
  \label{eq:defsigmahat}
  \hat\sigma(z) := \sqrt{\frac{4 m_\pi^2}{z}-1}  \,. 
\end{eqnarray}
Note that the point $s_c$ has been chosen close to the two-pion threshold such that the anomalous integral \eqref{eq:tanom} does not run over the high-energy region where the representations \eqref{tChPT1} and \eqref{tIAM1} become unreliable.

The value for the pion decay constant in the chiral limit $F_0$ is taken from the ratio $F_\pi/F_0=1.064(7)$, where
$F_\pi = 92.28(9)\,$MeV is the pion decay constant at the physical point. 
In the present work the low-energy constant $\bar{l}$ is adjusted such that the pion p-wave phase shifts from \eqref{tIAM1} agrees with that from \cite{Garcia-Martin:2011iqs} at the point $s_c$; see the discussion preceding (\ref{eq:tpsIAM}) below. In practice we use $\bar{l}=6.099$. This compares well with previous choices. In the original paper \cite{Dax:2018rvs}, the low-energy constant $\bar l = 5.73(8)$ has been adjusted such as to reproduce the position of the
pole of the $\rho$-meson resonance on the second Riemann sheet. In \cite{Junker:2019vvy}
$\bar l = 6.47$ had been used. 

Finally, we provide the anomalous piece of the TFFs:
\begin{eqnarray}
  G^{\rm anom}_m(q^2) &=& \frac{1}{12\pi} \, \int\limits_{0}^1 \text{d}x \, \frac{\text{d}s''(x)}{\text{d}x} \,
                          \frac{1}{s''(x)-q^2}
                          \nonumber \\ && \times
  \frac{f_m(s''(x)) \, s''(x) \, F^{V}_\pi(s''(x))}{-4\,  (-\lambda(s''(x),m_{\Delta}^2,m_n^2))^{3/2}}  \,,
  \label{eq:Fanom-unsub}
\end{eqnarray} 
this time with the straight-line path
\begin{eqnarray}
  \label{eq:defsx2}
  s''(x) := (1-x) s_+ + x\, s_Y
\end{eqnarray}
that connects the branch point (\ref{eq:defspconcr}) of the logarithm in (\ref{eq:Klog}) to the point $s_Y$ on the unitarity cut. The choice of this integration path simplifies the calculations as explained in the next subsection.

\subsection{Further modifications of the integration contour}
\label{sec:anom-num}

The standard path of the branch cuts --- from $s_+$ to the two-pion threshold $s_{2\pi}=4m_\pi^2$ and then along the 
real axis to $+\infty$ --- involves an integrand that is singular at the decay 
threshold 
$s_{dt}=(m_\Delta-m_N)^2$. Even though this singularity 
is integrable with the epsilon prescription $m_\Delta^2 \to m_\Delta^2+i\epsilon$ \cite{Bronzan:1963mby}, it is 
numerically easier to avoid this problem. 

Let the original integrals that we want to calculate be given by 
\begin{eqnarray}
  \label{eq:original}
  F(s) := \int\limits_{{\cal C}_{+,2\pi}} \! dz \, \frac{J_1(z)}{z-s-i\epsilon'} 
  + \int\limits_{4m_\pi^2}^\infty \! ds' \, \frac{J_2(s')}{s'-s-i\epsilon'} 
\end{eqnarray}
with the path ${\cal C}_{+,2\pi}$ connecting $s_+$ to the two-pion threshold. 
Consider a path along the closed triangle formed by $s_+$, 
the two-pion threshold $s_{2\pi}$ and an arbitrary point $s_{\rm c}$ on the real axis above the decay threshold $s_{dt}$. 
An integral over a function along this closed path vanishes, if this function is analytic inside of this triangle. 
This is the case for integrands $I(z)$ of the type
\begin{eqnarray}
  \label{eq:type}
  I(z) = \frac{J_1(z)}{z-s-i\epsilon'} \sim \frac{1}{[-\lambda(z,m_\Delta^2+i\epsilon,m_N^2)]^{3/2}} \, \frac{1}{z-s-i\epsilon'} \,.
\end{eqnarray}
Here $s$ lies on the real axis and the square root function is defined with a cut along the negative real axis. With the 
$\epsilon$ prescription for the mass of the unstable $\Delta$, the function $-\lambda(z,m_\Delta^2+i\epsilon,m_N^2)$ adopts 
negative real values slightly above the real axis (with real parts below $s_{dt}$ or above $s_{st}$). 

Thus instead of \eqref{eq:original} we can write 
\begin{eqnarray}
  \label{eq:better}
  F(s) = \int\limits_{{\cal C}_{+,{\rm c}}} \! dz \, \frac{J_1(z)}{z-s-i\epsilon'} 
  + \int\limits_{4m_\pi^2}^{s_{\rm c}} \! ds' \, \frac{J_2(s')-J_1(s')}{s'-s-i\epsilon'} 
  + \int\limits_{s_{\rm c}}^\infty \! ds' \, \frac{J_2(s')}{s'-s-i\epsilon'} 
\end{eqnarray}
with the path ${\cal C}_{+,{\rm c}}$ connecting $s_+$ to $s_{\rm c}$. What we have used to obtain \eqref{eq:better} is 
\begin{eqnarray}
  \label{eq:zeroclosedpath}
  0 = \int\limits_{{\cal C}_{+,2\pi}} \! dz \, \frac{J_1(z)}{z-s-i\epsilon'} 
  + \int\limits_{4m_\pi^2}^{s_{\rm c}} \! ds' \, \frac{J_1(s')}{s'-s-i\epsilon'} 
  - \int\limits_{{\cal C}_{+,{\rm c}}} \! dz \, \frac{J_1(z)}{z-s-i\epsilon'} \,.
\end{eqnarray}

The difference $J_2(s')-J_1(s')$ in \eqref{eq:better} involves just the standard 
logarithm/arctan without the extra term $\sim 2\pi i$. This difference diverges neither at 
the decay threshold $s_{dt}$ nor at the two-pion threshold $s_{2\pi}$. To be slightly more specific: 
\begin{eqnarray}
  J_2(s')-J_1(s') &\sim& \log \qquad \mbox{for} \quad s_{2\pi} < s' < s_{dt} \,, \nonumber \\
  J_2(s')-J_1(s') &\sim& \arctan \qquad \mbox{for} \quad s_{dt} < s' < s_{\rm c} \,, \nonumber \\
  J_2(s') &\sim& \arctan + \pi \qquad \mbox{for} \quad s_{\rm c} < s' < s_Y \,, \nonumber \\
  J_2(s') &\sim& \arctan  \qquad \mbox{for} \quad s_Y < s' < s_{st}\,, \nonumber \\
  J_2(s') &\sim& \log  \qquad \mbox{for} \quad s_{st} < s' \,.
  \label{eq:morespec}
\end{eqnarray}
Here $s_Y$ denotes the point where $Y(s)$ vanishes. Obviously, the simplest choice would be $s_{\rm c}=s_Y$. Then we could 
use in \eqref{eq:better} the standard log or standard arctan along the whole real axis. But we will argue 
below that this is {\em not} a good choice for $s_{\rm c}$. 

For the calculation of \eqref{eq:better} the only numerically problematic point is at $s=s_{\rm c}$. 
Since this point is arbitrary, the resulting function $F(s)$ must be 
smooth at this point. Schematically we can rewrite each of the integrals of \eqref{eq:better} into 
\begin{eqnarray}
  \label{eq:rewritesmooth}
  \int \! dz \, \frac{J_{\ldots}(z)}{z-s-i\epsilon'} = \int \! dz \, \frac{J_{\ldots}(z)-J_{\ldots}(s)}{z-s-i\epsilon'} 
  + J_{\ldots}(s) \int \! dz \, \frac{1}{z-s-i\epsilon'} \,.
\end{eqnarray}
Here $J_{\ldots}$ denotes $J_1$, $J_2$ or $J_1-J_2$. The first term on the right hand side of \eqref{eq:rewritesmooth} 
is smooth for any value of $s$ if 
\begin{eqnarray}
  \label{eq:smoothness}
  J_{\ldots}(z)-J_{\ldots}(s) \sim z-s \qquad \mbox{for} \quad z \to s \,.
\end{eqnarray}
The second term is proportional to $J_{\ldots}(s) \, \log(s_{\rm c}-s)$. Such a term diverges logarithmically for $s \to s_{\rm c}$. 
If one takes all integrals of \eqref{eq:better} together, one obtains for the potentially divergent terms a sum proportional to 
\begin{eqnarray}
  \label{eq:sumlogs}
  J_1(s) \, \log(s_{\rm c}-s) + \left(J_2(s)-J_1(s) \right) \, \log(s_{\rm c}-s) - J_2(s) \, \log(s_{\rm c}-s) = 0 \,.
\end{eqnarray}
Thus there is no numerical problem with \eqref{eq:better} if something like \eqref{eq:rewritesmooth} and \eqref{eq:sumlogs} 
is numerically implemented and {\em if} \eqref{eq:smoothness} is satisfied. 

To be more specific one needs in particular
\begin{eqnarray}
  \label{eq:smoothness2}
  J_1(z)-J_1(s_{\rm c}) \sim z-s_{\rm c} \qquad \mbox{for} \quad z \to s_{\rm c} \,.
\end{eqnarray}
Here $z$ is a complex number on the line that connects $s_+$ with $s_{\rm c}$. 
All this resembles to some extent the discussion for the two-pion threshold in \cite{Junker:2019vvy}. We use 
in practice two versions for the two-pion scattering amplitude. One along the real axis based on the measured phase shift, 
$t_{\rm ps}$, and the other, $t_{\rm IAM}$, employed in the complex 
plane for the definition of $J_1$. Those two versions must agree at the connection $s=s_{\rm c}$ to make $F(s)$ smooth, i.e.\ to 
ensure that \eqref{eq:smoothness2} is satisfied. The latter is necessary, otherwise the cancellation \eqref{eq:sumlogs} does 
not happen. The crucial point is that $J_1(z)$ is obtained from $t_{\rm IAM}$ 
while $J_1(s_{\rm c})$ is obtained from $t_{\rm ps}$.  
What needs to be done in practice is to readjust the low-energy constant that appears in $t_{\rm IAM}$ such that 
\begin{eqnarray}
  \label{eq:tpsIAM}
  {\rm Re}t^{-1}_{\rm ps}(s_{\rm c}) = {\rm Re}t^{-1}_{\rm IAM}(s_{\rm c}) 
\end{eqnarray}
holds.

As spelled out in \cite{Junker:2019vvy}, we trust the complex-plane two-pion amplitude $t_{\rm IAM}$ in the low-energy region. 
Thus $s_{\rm c}$ should not be chosen too large. More generally, the whole path ${\cal C}_{+,{\rm c}}$ should lie in the 
low-energy region. In practice, the branch point of the anomalous branch cut lies at $s_+ \approx (0.0458-0.0827\,i)\,$GeV$^2$. 
This is in size comparable to the two-pion threshold $s_{2\pi} \approx 0.0779\,$GeV$^2$. 
On the other hand, the point where $Y(s)$ vanishes lies at $s_Y\approx 0.675\,$GeV$^2$, which is not very small. 
Thus we should choose
$s_{\rm c}$ larger than the decay threshold $s_{dt} \approx 0.0858 \,$GeV$^2$, but much smaller than $s_Y$. 
A convenient choice 
should be $s_{\rm c}= 5 m_\pi^2 \approx 0.0974\,$GeV$^2$.

\section{Large-$N_c$ relations}
\label{sec:largeNc}

This appendix has the purpose to provide an educated guess for the relative sign between the $\Delta$-nucleon-pion coupling constant $h_A$ and the TFFs provided by the experimental groups. This educated guess --- based on calculations for a large number of quark colors $N_c$ --- will be further supported by a direct comparison to the experimental results exploring both signs of $h_A$. Still it is encouraging to have additional theoretical support for our final choice. 

We extend the chiral Lagrangian such that it provides the interactions of baryons with Goldstone bosons and photons 
up to (including) next-to-leading order \cite{Holmberg:2018dtv}:
\begin{eqnarray}
  && \frac{D}{2} \, {\rm tr}(\bar B \, \gamma^\mu \, \gamma_5 \, \{u_\mu,B\}) 
  + \frac{F}{2} \, {\rm tr}(\bar B \, \gamma^\mu \, \gamma_5 \, [u_\mu,B])  \nonumber \\[0.7em]
  && {} + \frac{h_A}{2\sqrt{2}} \, 
  \left(\epsilon_{ade} \, \bar B^e_c \, (u^\mu)^d_b \, T_\mu^{abc} 
    + \epsilon^{ade} \, \bar T^\mu_{abc} \, (u_\mu)^b_d \, B^c_e \right) 
  \nonumber \\[0.7em]
  && {}+
  b_{M,D} \, {\rm tr}(\bar{B}\{ f_+^{\mu\nu},\sigma_{\mu\nu} B\})
  +  b_{M,F} \, {\rm tr}(\bar{B}[ f_+^{\mu\nu},\sigma_{\mu\nu} B]) \nonumber \\[0.7em]
  && {} +i \, c_M \left( \epsilon_{ade} \, \bar B^e_c \, \gamma_\mu \gamma_5 (f_+^{\mu\nu})^d_b \, T_\nu^{abc} 
    - \epsilon^{ade} \, (\bar T_\nu)_{abc} \, \gamma_\mu \gamma_5 (f_+^{\mu\nu})_d^b \, B_e^c \right)  \,.
  \label{eq:baryonlagrNLO}
\end{eqnarray}
In the framework of \cite{Jenkins:1991es,Dashen:1993as,Jenkins:2002rj} the baryons are treated as quasi non-relativistic fields. 
Four-component spinors $B$ are projected on their two-component particle content $B_v$ where $v$ denotes the baryon velocity.
For simplicity we choose the rest frame $v=(1,0,0,0)$. Then the dominant components of $T^\mu$ are the spatial components.
They satisfy 
\begin{eqnarray}
  \label{eq:Tvdotsigma}
  T_v^k \sigma^k = 0
\end{eqnarray}
with the Pauli matrices $\sigma^k$. The dominant parts of the spinor matrices are given by 
\begin{eqnarray}
  \gamma^k \gamma_5 \to \sigma^k \,, \nonumber \\
  \sigma^{i j} \to \epsilon^{ijk} \sigma^k \,, \nonumber \\
  \gamma^0 \gamma_5 \approx 0 \,, \nonumber \\
  \sigma^{0 j} \approx 0 \,.
  \label{eq:reduce-non-rel}
\end{eqnarray}

Essentially one can treat the axial-vector field $u_k$ and the magnetic field $\epsilon^{ijk} f^+_{ij}$ on equal footing. 
Using \eqref{eq:reduce-non-rel}, the interaction terms \eqref{eq:baryonlagrNLO} can be approximated in the following way:
\begin{eqnarray}
  && \frac{D}{2} \, {\rm tr}(B_v^\dagger \, \sigma^k \, \{u_k,B_v\}) 
  + \frac{F}{2} \, {\rm tr}(B_v^\dagger \, \sigma^k \, [u_k,B_v])  \nonumber \\[0.7em]
  && {} + \frac{h_A}{2\sqrt{2}} \, 
  \left(\epsilon_{ade} \, (B_v^\dagger)^e_c \, (u_k)^d_b \, T_v^{k,abc} 
    + \epsilon^{ade} \, T^{k \dagger}_{v,abc} \, (u_k)^b_d \, (B_v)^c_e \right) 
  \nonumber \\[0.7em]
  && {}+
  b_{M,D} \, {\rm tr}(B^\dagger_v \{ f_+^{ij},\epsilon^{ijk} \sigma^k B_v \})
  +  b_{M,F} \, {\rm tr}(B_v^\dagger [ f_+^{ij},\epsilon^{ijk} \sigma^k B_v ]) \nonumber \\[0.7em]
  && {} +i \, c_M \left( \epsilon_{ade} \, (B_v^\dagger)^e_c \, \sigma^j \, (f_+^{jk})^d_b \, T_{v}^{k,abc} 
    - \epsilon^{ade} \, T^{k\dagger}_{v,abc} \, \sigma^j \, (f_+^{j k})_d^b \, (B_v)_e^c \right)  \,.
  \label{eq:baryonlagr22}
\end{eqnarray}
Replacing $u_k$ by $\epsilon^{ijk} f^+_{ij}$, one proceeds from the $D$ ($F$) term to the $b_{M,D}$ ($b_{M,F}$) term. 

The relation between the $h_A$ term and the $c_M$ term is less obvious. Here the constraint \eqref{eq:Tvdotsigma} comes into 
play. We first rewrite parts of the $c_M$ terms (flavor indices are suppressed; ordering matters for the spinor structure):
\begin{eqnarray}
  \label{eq:long-rewr}
  \sigma^j \, f_+^{jk} \, T_{v}^{k} &=& \frac12 \left( \sigma^j \, T_{v}^{k} - \sigma^k \, T_{v}^{j} \right) f_+^{jk} 
  = \frac12 \, \sigma^l \, T_{v}^{m} \left( \delta^{lj} \, \delta^{mk} -  \delta^{lk} \, \delta^{mj} \right) f_+^{jk} \nonumber \\
  &=& \frac12 \, \sigma^l \, T_{v}^{m} \, \epsilon^{lmn} \epsilon^{jkn} \, f_+^{jk} 
  = - \frac14 i \, [\sigma^m,\sigma^n] \, T_{v}^{m} \, \epsilon^{jkn} \, f_+^{jk}  
  = - \frac14 i \, \sigma^m \, \sigma^n \, T_{v}^{m} \, \epsilon^{jkn} \, f_+^{jk}  \nonumber \\
  &=& -\frac14 i \, \{ \sigma^m , \sigma^n \} \, T_{v}^{m} \, \epsilon^{jkn} \, f_+^{jk}  
  = -\frac14 i \, 2 \delta^{nm}\, T_{v}^{m} \, \epsilon^{jkn} \, f_+^{jk}  
  = -\frac12 i \, T_{v}^{m} \, \epsilon^{jkm} \, f_+^{jk}  \nonumber \\
  &=& -\frac12 i \, T_{v}^{k} \, \epsilon^{ijk} \, f_+^{ij} \,;
\end{eqnarray}
\begin{eqnarray}
  \label{eq:long-rewr2}
  T^{k\dagger}_{v} \, \sigma^j \, f_+^{jk} &=& \frac12 \left( T^{k\dagger}_{v} \, \sigma^j - T^{j\dagger}_{v} \, \sigma^k \right) f_+^{jk} 
  = \frac12 \, T^{m\dagger}_{v} \, \sigma^l \left( \delta^{lj} \, \delta^{mk} -  \delta^{lk} \, \delta^{mj} \right) f_+^{jk} \nonumber \\
  &=& \frac12 \, T^{m\dagger}_{v} \, \sigma^l \, \epsilon^{lmn} \epsilon^{jkn} \, f_+^{jk} 
  = - \frac14 i \, T_{v}^{m\dagger} \, [\sigma^m,\sigma^n] \, \epsilon^{jkn} \, f_+^{jk}  
  = \frac14 i \, T_{v}^{m\dagger} \, \sigma^n \, \sigma^m \, \epsilon^{jkn} \, f_+^{jk}  \nonumber \\
  &=& \frac14 i \, T_{v}^{m\dagger} \, \{ \sigma^n , \sigma^m \} \, \epsilon^{jkn} \, f_+^{jk}  
  = \frac14 i \, T_{v}^{m\dagger} \, 2 \delta^{nm}\, \epsilon^{jkn} \, f_+^{jk}  
  = \frac12 i \, T_{v}^{m\dagger} \, \epsilon^{jkm} \, f_+^{jk}  \nonumber \\
  &=& \frac12 i \, T_{v}^{k\dagger} \, \epsilon^{ijk} \, f_+^{ij} \,.
\end{eqnarray}
Thus we can rewrite \eqref{eq:baryonlagr22} into 
\begin{eqnarray}
  && \frac{D}{2} \, {\rm tr}(B_v^\dagger \, \sigma^k \, \{u_k,B_v\}) 
  + \frac{F}{2} \, {\rm tr}(B_v^\dagger \, \sigma^k \, [u_k,B_v])  \nonumber \\[0.7em]
  && {}+
  b_{M,D} \, {\rm tr}(B^\dagger_v \, \sigma^k \, \{ \epsilon^{ijk} \, f_+^{ij},B_v \})
  +  b_{M,F} \, {\rm tr}(B_v^\dagger \, \sigma^k \, [ \epsilon^{ijk} \, f_+^{ij},B_v ]) \nonumber \\[0.7em]
  && {} + \frac{h_A}{2\sqrt{2}} \, 
  \left(\epsilon_{ade} \, (B_v^\dagger)^e_c \, T_v^{k,abc} \, (u_k)^d_b 
    + \epsilon^{ade} \, T^{k \dagger}_{v,abc} \, (u_k)^b_d \, (B_v)^c_e \right) 
  \nonumber \\[0.7em]
  && {} +\frac12 c_M \left( \epsilon_{ade} \, (B_v^\dagger)^e_c \, T_{v}^{k,abc} \, \epsilon^{ijk} \, (f_+^{ij})^d_b 
    + \epsilon^{ade} \, T_{v,abc}^{k\dagger} \, \epsilon^{ijk} \, (f_+^{ij})_d^b \, (B_v)_e^c \right)  \,.
  \label{eq:baryonlagr23}
\end{eqnarray}
Therefore the large-$N_c$ formalism relates $c_M/\sqrt{2}$ 
in the same way to $b_{M,D}$ and $b_{M,F}$ as it relates $h_A$ to $D$ and $F$. 
In particular, if conventions are chosen such that $h_A$ has the same positive sign as $D$ and $F$, then $c_M$ must have the same sign as 
$b_{M,D}$ and $b_{M,F}$. 

At next-to-leading order of baryon $\chi$PT, i.e.\ using \eqref{eq:baryonlagrNLO}, 
the anomalous magnetic moments of proton and neutron are 
determined by $b_{M,D}$ and $b_{M,F}$. The TFF $F_1$ is determined by $c_M$. 

It turns out that the respective signs of the magnetic moments of proton and neutron lead to positive values for 
$b_{M,D}$ and $b_{M,F}$ \cite{Kubis:2000aa,KubisPhD}. Then the large-$N_c$ considerations lead to a positive value for $c_M$. 
In turn, this leads to a positive value for $F_1(0)$. 
Consequently, 
we obtain a negative value for $G_{-1}(0)$ from \eqref{eq:F-1def} and a positive value for $G_{+1}(0)$ from \eqref{eq:F+1def}. 
Therefore the combination  
$G_{-1}(0) - 3 G_{+1}(0)$ is negative. On the other hand, the usual convention for the magnetic Jones-Scadron TFF is such that $G^*_M(0)$ is positive \cite{Pascalutsa:2006up}. To obtain such a positive magnetic transition moment requires an overall negative sign in 
\eqref{eq:Gs-jones-scadron} and therefore \eqref{eq:zeta-choice}. 

Finally, we note that these results are in complete agreement with the calculations of \cite{Junker:2019vvy}. There is 
a relative minus sign in the flavor factors between the transitions $\Sigma^*$-$\Lambda$ and $\Delta$-$N$ as can be read off 
from table 3 of \cite{Holmberg:2018dtv}. Thus the positive value of $G_{-1}(0)$ for the hyperon case in 
table I of \cite{Junker:2019vvy} translates to a negative value for the $\Delta$-nucleon case. 
The negative value of $G_{+1}(0)$ for the hyperon case translates to a positive value for the $\Delta$-nucleon case. All this agrees
with the analysis of the previous paragraph. 

To summarize the large-$N_c$ prediction: A positive value for $h_A$ leads to a negative value of $\zeta$ in \eqref{eq:zeta-choice} and vice versa. A positive value for $\zeta$ would lead to a negative value for $h_A$. 

In practice, we demand \eqref{eq:zeta-choice}. Then we expect to find a positive value of $h_A$. But we will explore both options by a comparison of our subtracted dispersion relation (\ref{eq:dispbasic}) to the experimental results for the $q^2$ dependence of $G^*_M(q^2)$.

\section{Contact terms from chiral perturbation theory}
\label{sec:contact-ChPT}

Our subtracted dispersion relations (\ref{eq:dispbasic}) and (\ref{eq:tmandel}) contain undetermined subtraction constants. They parametrize our limited knowledge of the high-energy region. The same statement holds for the low-energy constants of effective field theories. If we insist that all pion-baryon scattering amplitudes are obtained from $\chi$PT (plus pion rescattering) then one can relate the polynomials $P_m$ to contact terms emerging at various orders of the chiral power counting. 

On the other hand, it has been found in \cite{Leupold:2017ngs} for the nucleon isovector FFs that the subtraction constants do not entirely agree with low-order estimates from $\chi$PT. We found the same for our case at hand. It is better to determine the subtraction constants from data. 

For completeness, however, we will present the results for the constants $P_m$ that are obtained from an NLO calculation. The first step is to specify the three-point interactions. Modifications can be compensated by changes of the contact terms (four-point interactions) \cite{Pascalutsa:2000kd,Granados:2017cib,Leupold:2017ngs}. 

For our purposes the interaction term proportional to
$H_A$ effectively reduces to
\begin{eqnarray}
  \label{eq:HAeff}
  + \frac{H_A}{2 m_R \, F_\pi} \, \epsilon_{\mu\nu\alpha\beta} \, 
  \bar T^\mu_{abc} \, \partial^\nu (T^\alpha)^{abd} \, \partial^\beta\Phi^c_d \,
\end{eqnarray}
where the resonance mass $m_R$ corresponds in this case to $m_\Delta$.
Working with relativistic spin-3/2 Rarita-Schwinger fields is plagued by ambiguities how to deal with the spurious spin-1/2 
components. In the present context the interaction term $\sim h_A$ causes not only the proper exchange of spin-3/2 resonances, 
but induces an additional contact interaction. This unphysical contribution can be avoided by constructing interaction terms 
according to the Pascalutsa description \cite{Pascalutsa:1999zz,Pascalutsa:2005nd,Pascalutsa:2006up,Ledwig:2011cx}. 
It boils down to the replacement
\begin{eqnarray}
  \label{eq:replace}
  T^\mu \to -\frac{1}{m_\Delta} \, \epsilon^{\nu\mu\alpha\beta} \, \gamma_5 \, \gamma_\nu \, \partial_\alpha T_\beta.
\end{eqnarray}
Strictly speaking this procedure induces an explicit flavor breaking, but such effects 
are anyway beyond leading order.
The $H_A$ term of (\ref{eq:HAeff}) is already constructed such that only the spin-3/2 components contribute.

We will explore both the standard interaction term $\sim h_A$ from (\ref{eq:baryonlagr}) and the corresponding one obtained 
by (\ref{eq:replace}). As already discussed in \cite{Junker:2019vvy},  differences can be accounted for by contact interactions 
of the chiral Lagrangian at NLO and beyond. 

The explicit expressions for the polynomial terms are
\begin{eqnarray}
  P_{+1} & = & -\frac{g_A h_{A}}{4 \sqrt{6} F_{\pi }^2} 2
  - \frac{5h_{A} H_{A}}{12 \sqrt{6} F_{\pi }^2} \, \frac{5 \, (m_{\Delta}+m_n)}{6m_{\Delta}}   \,, \nonumber \\
  P_{-1} & = & -\frac{g_A h_{A}}{4 \sqrt{6} F_{\pi }^2} \frac{2 \, (m_{\Delta}-m_n-m_p)}{m_{\Delta}} \nonumber \\
               && {} - \frac{5h_{A} H_{A}}{12 \sqrt{6} F_{\pi }^2}
               \, \frac{s-2 m_\pi^2 - (m_{\Delta}+m_n)(6m_{\Delta}-m_n)}{6 m_{\Delta}^2}
               \nonumber \\
         &\approx& -\frac{g_A h_{A}}{4 \sqrt{6} F_{\pi }^2} \frac{2 \, (m_{\Delta}-m_n-m_p)}{m_{\Delta}} \nonumber \\
               && {} - \frac{5h_{A} H_{A}}{12 \sqrt{6} F_{\pi }^2} \,
               \frac{-(m_{\Delta}+m_n)(6m_{\Delta}-m_n)}{6 m_{\Delta}^2}
               \,, \nonumber \\
  P_{0} & = & -\frac{g_A h_{A}}{4 \sqrt{6} F_{\pi }^2}
  - \frac{5h_{A} H_{A}}{12 \sqrt{6} F_{\pi }^2}\, \frac{3 m_{\Delta}-m_n}{6 m_{\Delta} }  \,.
  \label{eq:P-pw}  
\end{eqnarray}
For $P_{-1}$ we dropped terms which are suppressed by two orders in the chiral counting. 

The $\Delta\bar{n}\pi^+\pi^-$ contact diagram produces the following polynomials:
\begin{eqnarray}
P^{\rm NLO\,\chi PT}_{+1}&=&-c_F \frac{m_{\Delta}+m_n}{\sqrt{3} F_\pi^2}   \,, \nonumber \\
P^{\rm NLO\,\chi PT}_{0}&=&-c_F \frac{m_{\Delta}}{\sqrt{3} F_\pi^2}  \,,  \nonumber \\
P^{\rm NLO\,\chi PT}_{-1}&=&c_F \frac{m_n(m_{\Delta}+m_n)-s}{\sqrt{3} F_\pi^2 \, m_{\Delta}} \nonumber \nonumber \\ &\approx &
 c_F \frac{m_n(m_{\Delta}+m_n)}{\sqrt{3} F_\pi^2\, m_{\Delta}} \,.
\end{eqnarray}

The amplitudes \eqref{eq:feynlo+nlo} become slightly different when the Pascalutsa prescription is used:
new contact terms pop up but the pole terms and therefore \eqref{eq:K-pw} are not affected. In particular we have:
\begin{equation}
\begin{split}
    P^{P}_{+1}&= P_{+1}  \\
    & {}-\frac{5h_A H_A}{36\sqrt{6}F_\pi^2m_{\Delta}^2}((m_n+m_{\Delta})(2m_{\Delta}+3m_n)-3s)   \,, \\
    P^{P}_{0} &= P_{0}
    +\frac{5h_A H_A}{36\sqrt{6}F_\pi^2}   \,, \\
    P^{P}_{-1}&=P_{-1}  \\
    & {}+\frac{5h_A H_A}{36\sqrt{6}F_\pi^2m_{\Delta}^2}((m_n+m_{\Delta})(3m_{\Delta}+2m_n)-2s).
\end{split}
\end{equation} 
As expected the proton exchange diagrams do not get any contribution since in that case the only $\Delta$ baryon is the external one, which is on-shell.

\bibliography{lit}{}
\bibliographystyle{apsrev4-1}
\end{document}